\documentclass[12pt]{article}
\usepackage[left=1in,top=1in,right=1in,bottom=1in]{geometry}
\usepackage{amsmath}
\usepackage{graphicx,amsmath,amssymb,fullpage,amsfonts,bbm,bbold}
\usepackage{amsbsy}
\usepackage{amsthm}
\usepackage{enumerate}
\usepackage{epsfig}
\usepackage{color}
\usepackage{url}
\usepackage[round]{natbib}
\usepackage{setspace}
\usepackage{bbm}
\usepackage[multiple]{footmisc}

\makeatletter
\usepackage{booktabs}

 \usepackage{setspace}
\newcommand{\E}{\mbox{E}}

\renewcommand\section{\@startsection {section}{1}{\z@}%\
                                  {-3.5ex \@plus -1ex \@minus -.2ex}%\
                                  {2.3ex \@plus.2ex}%\
                                  {\centering\scshape}}

\renewcommand\subsection{\@startsection {subsection}{1}{\z@}%\
                                  {-3.5ex \@plus -1ex \@minus -.2ex}%\
                                  {2.3ex \@plus.2ex}%\
                                  {\bf\normalsize}}
                                  
\renewcommand\subsubsection{\@startsection {subsubsection}{1}{\z@}%\
                                  {-3.5ex \@plus -1ex \@minus -.2ex}%\
                                  {2.3ex \@plus.2ex}%\
                                  {\normalsize}}

\date{}
\begin{document}
\begin{center}
\begin{doublespace}
{\Large\bf A Bayesian hierarchical model for inferring player strategy types in a number guessing game}
\end{doublespace}

\vspace{.3in}
\begin{singlespace}
{\small
{\sc P.~Richard Hahn and Indranil Goswami} \\ \smallskip
{\it Booth School of Business, University of Chicago \\
Chicago, Illinois~60637-1656, U.S.A.} \\ 
{richard.hahn@chicagobooth.edu} \medskip
\vspace{0.1in} \\
{\sc Carl Mela} \\ \smallskip
{\it T. Austin Finch Foundation Professor of Business Administration\\
Fuqua School of Business, Duke University \\
Durham, North Carolina~27708-0251, U.S.A.}} \\ \medskip

\end{singlespace}
\end{center}
\section*{Abstract}
\begin{singlespace}
This paper presents an in-depth statistical analysis of an experiment designed to measure the extent to which  players in a simple game behave according to a popular behavioral economic model.  The $p$-beauty contest is a multi-player number guessing game that has been widely used to study strategic behavior.  This paper describes beauty contest experiments for an audience of data analysts, with a special focus on a class of models for game play called $k$-step thinking models, which allow each player in the game to employ an idiosyncratic strategy.  We fit a Bayesian statistical model to estimate the proportion of our player population whose game play is compatible with a $k$-step thinking model. Our findings put this number at approximately 25\%.\\

\noindent {\small \textit{Key words:} behavioral game theory, hierarchical modeling, partial identification.}\\

\vspace{1.25in}
\noindent {\small \textit{Acknowledgements:} We would like to thank the University of Rochester, the California Institute of Technology, and Oleg Urminsky, Abigail Sussman, and George Wu for helpful discussions, Ben Gillen, Colin Camerer, Charles Plott, and two anonymous referees for helpful comments, and Teck-Hua Ho for comments and sharing data. Thanks also to Jared Murray, Dan Merl and Kristian Lum for helpful feedback.  The first author would like to thank the Booth School of Business for support.}

\end{singlespace}
\newpage
\section{Introduction}
Game theory provides a formal language for describing situations where outcomes are determined by the behavior of multiple individuals, all acting in their own interests.  By ``solving puzzles about how idealized players will behave"  \citep{camerer2003behavioural}, game theoretic models admit rigorous analysis and furnish sharp predictions, making them an important tool for characterizing agent interactions in a variety of areas, from biological, to social, to industrial.  

However, in some contexts---such as first-time interactions---the predictions from idealized game theoretic models fail to characterize observed game play. Mindful of this discrepancy, {\em behavioral game theory} takes up the challenge of describing how people actually play games (in the broad sense). Behavioral game theorists use ``experimental evidence to inform mathematical models of cognitive limits, learning rules and social utility" \citep{camerer2003behavioural}, combining the rigor and precision of mathematical game theory with the experimental methods of cognitive science.  Behavioral game theory is an active and rapidly expanding field (e.g., \cite{chong2014generalized}), and to even outline its recent developments would require far more space than we have here.  For a book length introduction see \citep{camererbook}.  Suffice it to say that improved understanding of how people reason strategically would be useful in a broad range of settings, from product pricing to battlefield decision-making. 

This paper analyzes game play data from a $p$-beauty contest, a classic game from the behavioral game theory literature.  A $p$-beauty contest, like the better-known prisoner's dilemma, is an easy-to-describe game that serves as an ideal instrument for eliciting and observing strategic behavior. Understanding how people approach simple multi-player games like beauty contests is an important step towards characterizing the ways that people tend to reason about the reasoning of others.  

More specifically, we fit a Bayesian hierarchical model to data from a web experiment where subjects play against one another in several $p$-beauty contest.  The goal of our analysis is to estimate the proportion of our player population whose observed game play is consistent with prevailing theories of strategic reasoning, while explicitly allowing for individual-level strategic heterogeneity. We intend that our case study may serve to introduce the wider community of applied statisticians to the data-rich field of behavioral game theory. At the same time, our new statistical approach to $p$-beauty data provides insights into player behavior that should be of interest to behavioral economists.

The balance of the introduction describes the beauty contest game and also presents a flexible ``$k$-step thinking" model for beauty contest data.  This model generalizes previously-studied models called cognitive hierarchy models and level-$k$ models.  An abridged literature review---providing pointers to the vast literature on beauty contest experiments, cognitive hierarchy and level-$k$ models---can be found in Section \ref{litreview}. The remainder of the paper is devoted to describing our experiment, the data we collected, the hierarchical model we use to analyze that data, and our empirical findings.

%Section five concludes by outlining the subtle but important ways our approach differs from standard (fully-identified) Bayesian estimation and testing; this section provides a detailed account of how the specific modeling goals of behavioral game theory motivated our approach.

\subsection{Beauty contest games}\label{beauty contest}
\begin{quote}
{\em
...professional investment may be likened to those newspaper competitions in which the competitors have to pick out the six prettiest faces from a hundred photographs, the prize being awarded to the competitor whose choice most nearly corresponds to the average preferences of the competitors as a whole...It is not a case of choosing those which, to the best of one's judgment, are really the prettiest, nor even those which average opinion genuinely thinks the prettiest. We have reached the third degree where we devote our intelligences to anticipating what average opinion expects the average opinion to be. And there are some, I believe, who practice the fourth, fifth and higher degrees.}
\end{quote}
\hspace{3in}---John Maynard Keynes \citep{keynes1936general}\\

Each player of a $p$-beauty contest is instructed to pick a number between, say, 0 and 100.  The game parameter $p$ is pre-determined, common, and known to all players. The player whose number is closest to $p$ times the average number across all players wins a fixed payout (winner-take-all).  Other variants of the game are possible, for example, with payoffs inversely proportional to the distance from the target of $p$ times the group average (continuous payout).

To briefly consider a concrete example, consider five participants playing a beauty contest with $p = 3/4$.  Suppose the players' numbers are 5, 15, 20, 40 and 50. To determine the winner, compute $p$ times the average of all the numbers:  $\frac{3}{4} \times (5+15+20+40+50)/5 = 19.5$.  The player who picked 20 wins, because among all the players' numbers, 20 is closest to 19.5.

The beauty contest is a symmetric game, meaning that all players have the same payoff function. When the total number of players, $n$, is more than several dozen, one's own play negligibly affects the overall group mean, so that the game is effectively a guessing game. In this paper, we restrict attention to $p \in [0,1]$.  Because every player is trying to undercut the group average by the fraction $p$, the Nash equilibrium strategy is zero, meaning that everyone playing zero is the mutually best response.

However, experiments consistently reveal that most people do not play the zero strategy.  On this point, the recent survey paper by \cite{crawford_review} notes (emphases added):
\begin{quote}
Although Nash equilibrium can be and has been viewed as a model of strategic thinking, experimental research shows with increasing clarity that subjects' initial responses to games often deviate systematically from equilibrium, and that the {\em deviations have a large structural component that can be modeled in a simple way}. Subjects' thinking tends to [favor] rules of thumb that {\em anchor beliefs in an instinctive reaction to the game and then adjust them via a small number of iterated best responses}.
\end{quote}
Our goal in this paper will be to estimate from data the prevalence of such rules of thumb.

\subsection{A structural model for beauty contest game play}\label{kstep}
Our model of strategy formation in beauty contest games supposes that player heterogeneity arises from two sources:  1) idiosyncratic beliefs about the strategies of others and 2) the number of iterations one proceeds, conditional on these beliefs, towards the Nash equilibrium.

To motivate this model, consider a player pondering her strategy\footnote{Assume she believes that the game has enough players so that sampling variation of the mean play of the group is negligible.}. Suppose she thinks that most people will not understand the game and will play randomly with mean $\mu^0$.  This is, in the words of \cite{crawford_review}, her ``instinctive reaction".  Accordingly, her optimal play is $p \mu^0$.  Upon further reflection, however, she realizes that some of her opponents may have come to the same conclusion and would also play $p \mu^0$.  This would mean that her optimal play would be $p$ times a weighted average of $\mu^0$ and $p\mu^0$, where the weights reflect her beliefs about the relative proportions of the first two types of players. We refer to a totally random player as a 0-step player, one who thinks a single extra step a 1-step player, and so forth.  

Let $\gamma_i$ denote the level thinking of player $i$.  An iterated reasoning strategy for player $i$ can be described via the following parameterization:
\begin{itemize}
\item a scalar parameter $\mu^0_i \in (0,100)$ that represents player $i$'s belief as to the mean play of the 0-step players, and 
\item a $(\gamma_i-1)$-by-$\gamma_i$ lower-triangular right stochastic matrix $\Omega_i$, which we call a {\em belief matrix} for reasons described below.
\end{itemize}
{\em Right stochastic} means that the row entries are positive numbers that sum to one. The bottom row of $\Omega_i$ represents the individual's beliefs about the relative proportions of the various strategy classes below her, while the above rows reflect her beliefs about the analogous beliefs of each of the corresponding strategy levels below $\gamma_i$.

Given these two parameters, one may compute the $(\mu_i^0, \Omega_i)$-optimal response from the following recursive formula:
\begin{equation}\label{recursion}
\begin{split}
\mu_i^1(p) &= p \mu_i^0,\\
\mu_i^h(p) &= p \sum_{j=1}^{h} \omega_{h,j}^i \mu_i^{j-1}(p),
\end{split}
\end{equation}
where $\omega_{h,j}^i$ denotes the $(h,j)$ entry of $\Omega_i$.  A player with parameters $(\Omega_i, \mu_i^0)$ has optimal response $\mu_i(p) \equiv \mu_i^{\gamma_i-1}(p)$; in other words, expression (\ref{recursion}) is applied until there are no more rows of $\Omega_i$.  We will refer to this model of strategy formation as a $k$-step thinking model.

Clearly, $\mu^0_i$ and the free elements of $\Omega_i$ are underdetermined given only a player's optimal response $\mu_i(p)$ for a finite set of values of $p$.  Earlier analyses of beauty contest data have addressed this difficulty by introducing additional restrictions on $\Omega_i$ and $\mu^0_i$, both in terms of the structure of $\Omega_i$ and also in terms of limiting heterogeneity by assuming some parameters are shared across players.  Our approach is distinct in that we not estimate $(\Omega_i, \mu^0_i)$.  Before sketching our approach, we briefly examine two previous models from the literature for comparison, to observe how their additional assumptions impose restrictions on $\Omega_i$ and $\mu_i^0$.

\subsubsection{Cognitive hierarchy models}
A cognitive hierarchy model \citep{CHC} stipulates that
\begin{itemize}
\item $\Omega_i =\Omega_j$ whenever $\gamma_i = \gamma_j$. That is, all players thinking the same number of steps have the same beliefs about their opponents (and their beliefs).
\item For $\gamma_j = \gamma_i - 1 > 2$,  $\Omega_j$ can be obtained from $\Omega_i$ by removing the bottom row and the right-most column and rescaling each row to maintain right stochasticity. Players thinking more steps ahead are able to accurately project themselves into the mindset of lower level players.
\end{itemize}
Thus, in a cognitive hierarchy model, the belief matrices of lower-level thinkers are nested inside the belief matrices of higher-level thinkers.  Specific cognitive hierarchy models may make additional assumptions.  Common ones include:
\begin{itemize}
\item $\mu^0_i$ is common across $i$; all players assume the same mean for the 0-step players,
\item $\mu^0$ is in fact the actual mean of the level-0 players,
\item  players' beliefs about the relative proportions of thinker types is accurate.
\end{itemize}
For example, if $\gamma_i = 3$, the CH-Poisson model of \cite{CHC} yields 
\begin{equation*}
\Omega_i = \begin{pmatrix}
\frac{f_2(0)}{f_2(0)+f_2(1)} = \frac{1}{3}&  \frac{f_2(1)}{f_2(0)+f_2(1)}=\frac{2}{3}  & 0 \\ 
\frac{f_2(0)}{f_2(0)+f_2(1)+f_2(2)} = \frac{1}{5} & \frac{f_2(1)}{f_2(0)+f_2(1)+f_2(2)} =\frac{2}{5}  & \frac{f_2(2)}{f_2(0)+f_2(1)+f_2(2)}=\frac{2}{5}
\end{pmatrix}
\end{equation*}
where $f_{\tau}(\cdot)$ denotes the Poisson probability mass function with parameter $\tau$ (set here to two).  In addition to determining the elements of $\Omega$ for each player type, the accuracy assumption (third bullet point just above) means that $\tau$ also governs the frequency of player types in the population.  

\subsubsection{Level-$k$ models}
A ``level-$k$" model (see \cite{nagel1995unraveling} and \cite{crawford2007}) stipulates that each player believes that all of her opponents are reasoning one step fewer than she is.  In our notation, if $\gamma_i = k$, $\Omega_i$ has $\omega_{g, g+1}^i =1$ and zeros elsewhere for $g = 1\dots k-1$.  When $k=3$ this gives
\begin{equation*}
\Omega_i = \begin{pmatrix}
0 & 1 & 0 \\
0 & 0 & 1
\end{pmatrix}.
\end{equation*}
While the belief matrix is assumed fixed for all player types, the distribution of players among these types is assumed to be general and can be estimated from data.

\subsubsection{Our approach: estimating the prevalence of $k$-step thinking}\label{prevalence}
Because the extreme flexibility of the heterogeneous $(\Omega_i, \mu^0_i)$ $k$-step thinking model precludes estimation of model parameters, our analysis instead centers around general properties that would be exhibited by any $k$-step thinking strategy, for {\em any} parameter values.  Specifically, any $k$-step thinking strategy will satisfy the following straightforward conditions:
\begin{enumerate}[i)]
\item at $p = 1$, if individual $i$ is a non-random player, her optimal play is $\mu_i^0$, 
\item at $p=0$, the optimal play for any non-random player is 0,
\item if $\gamma_i = k$, then her optimal play lies in the interval $(p^k\mu_i^0, p \mu_i^0)$ (this follows from the extreme cases of assuming that all players are ($k-1$)-step or 0-step players, respectively), and  
\item a $k$-step iterated reasoning strategy is a positive linear combination of the monomial terms $p,p^2,\dots,p^k$, so is strictly increasing in $p$ and convex on $[0,1]$.
\end{enumerate}
Our statistical analysis will assess the probability that these conditions are satisfied for a randomly selected player from the population, by fitting a purely descriptive spline-based model to each player's response data.  Because the descriptive model is not restricted to satisfy the criteria above, it allows that some players may not play a $k$-step thinking strategy of any form, without requiring that such a player is a 0-step thinker in the sense of playing entirely at random.  

Our approach differs from most previous statistical analyses of beauty contest data in that we aim neither to {\em estimate} a particular identified instantiation of a $k$-step model, nor to {\em test} if a given instantiation fits a particular data set well.  Instead, our approach puts player heterogeneity at its center and asks a related question:  what proportion of players' responses are consistent with some {\em implicit} $k$-step thinking model? Using this approach, we are able to determine approximately how common $k$-step thinking is, in a way that mitigates distortions due to potential model mis-specification.   

\subsection{Literature overview}\label{litreview}
To be concise, the  literature review below is abridged, geared primarily towards providing background for the present data analysis.  For a scholarly and comprehensive review, we recommend the survey article quoted above, \cite{crawford_review}.  Specifically, Section 3 of that paper reviews the beauty contest literature at length. In what follows, we quote directly from their definitive exposition in framing the contribution of this paper.

\subsubsection{Data and theories}
Among the first empirical studies of beauty contest experiments are \cite{nagel1995unraveling}, \cite{ho1998iterated}, and \cite{bosch2002one}.  The first two studies directed subjects to play the game multiple times sequentially, with the outcome revealed after each round (feedback), so only the first round of play can be regarded as an initial response.  Restricting attention to initial response data serves to isolate the role of strategic reasoning from the influence of adaptation to outcomes from previous rounds of play.  The \cite{bosch2002one} study is notable for being a large-scale newspaper experiment with several thousand respondents. The main message of these studies was two-fold.  First, ``[s]ubjects seldom made equilibrium guesses initially" and second, ``[s]ubjects' thinking is heterogeneous, so no model that imposes homogeneity will do justice to their behavior" \citep{crawford_review}.  A characteristic feature of the distribution of guesses in these experiments was local modes at approximately $50p$, $50p^2$ and $50p^3$ across different values of $p$.  This feature  was hypothesized to arise via players applying iterated best response $k$ times for differing values of $k$ (as in the structural model described above), anchoring on an initial guess of 50.  However, this interpretation is underpinned by the strong assumption that players share the same initial guess of 50. Also, ``subjects' initial responses were limited to one game. One observation yields very limited information about the rule a subject was following" \citep{crawford_review}.  Later papers collected multiple initial responses for differing values of $p$, without feedback, notably \cite{coricelli2009neural}, but none on the large scale of \cite{bosch2002one}.  In our study, we collected multiple initial responses (without feedback) for multiple values of $p$ per subject.  We recruit more participants than \cite{coricelli2009neural} (whose main focus was conducting fMRI experiments), and do not require subjects' initial guess to be shared.
 
 \subsubsection{Model fitting and evalution}
On the statistical front, a common approach to investigating iterated reasoning has been to specify a structural model, to propose an error model to capture moderate deviations from the defined structure, and to fit maximum-likelihood estimates to the model parameters.  Particularly popular are mixture models positing a discrete number of player types.  Influential papers taking this approach include \cite{Stahl-Wilson-1995} and \cite{costagomes} (although neither of these papers considers beauty contests specifically).  This approach has two limitations.  First, maximum likelihood estimation is known to suffer from over-fitting of the data, especially in latent variable models where the number of unknown parameters (player type indicators in this case) scales with the number of observations. This problem can be mitigated using penalized likelihood methods or, as we do here, Bayesian priors and/or hierarchical models.  In a similar vein, finite mixture models present the possibility of model mis-specfication in terms of the number of types allowed.  \cite{costagomes} is mindful of this possibility, but the hazard is underscored by the fact that they leave 33 of their 88 subjects unclassified.  In this paper we allow each subject to play their own distinct strategy, while using hierarchical priors to ``shrink"  strategies towards one another so as to prevent over-fitting.

A related, and more serious, drawback is that likelihood-based model selection is inherently sensitive to the specification of null hypotheses and inessential modeling decisions (such as error distributions). One approach to overcoming this difficulty has been to seek external validating evidence to ``test a model's specification and evaluate the credibility of its explanation of behavior" \cite{crawford_review}.  A promising method in this direction ``is to study cognition via measures that complement decisions, such as monitoring subjects' searches for hidden information or monitoring their neural activity" \citep{crawford_review}.  Techniques such as gaze-tracking \citep{wang2010pinocchio} or mouse-tracking \citep{costa2001cognition} have been used for this purpose. Functional magnetic resonance imaging (fMRI) studies have even been conducted on subjects as they play the game.  For example, \cite{coricelli2009neural} correlate estimated level thinking in a beauty contest with neural activity in the medial prefrontal cortex.  For additional studies taking similar auxiliary confirmatory evidence approaches, see again \citep{crawford_review}, specifically section 3.5, page 23.

In our study, we do not seek such external validating evidence, but neither do we test our model against a null model with a formal statistical test.  Instead, our approach is to use multiple values of $p$ as a confirmatory experiment.  Specifically, the $k$-step thinking hypothesis entails a specific prediction about how people ought to play as a function of $p$.  In a Popperian spirit \citep{gelman_shalizi}, we simply collect data and check---using a flexible statistical model---how often players' responses appear to satisfy that condition.

\section{Data collection}
\subsection{Estimation strategy}\label{estimation}
The primary focus of our analysis will be to assess if subjects' game responses are consistent with a $k$-step thinking strategy. As discussed in Section \ref{prevalence}, it is possible to address this question without having to reference the structural parameter $\Omega_i$.  Instead, we need only check that a player's strategy function satisfies two necessary (though not sufficient) conditions:  i) that it runs through the origin, and ii) that it is an increasing convex function of the game parameter $p$.  We will refer to these two conditions as the $k$-step compatibility criteria.  

Because we do not {\em impose} this characteristic $k$-step shape during estimation, data which nonetheless exhibit this shape provide an estimate of (an upper bound on) the proportion of players who $k$-step reason.  Critically, this approach demands collecting data for multiple values of the game parameter $p$ per subject without feedback.  We now describe our procedure for collecting such data.

\subsection{Study protocol}
Our data were collected from Amazon's Mechanical Turk, a web interface that allows anonymous compensation to game participants.  See \cite{turk} for an introduction to the use of Amazon Mechanical Turk in social science research.

\begin{figure}[h!]
\begin{center}
\includegraphics[width=4.5in]{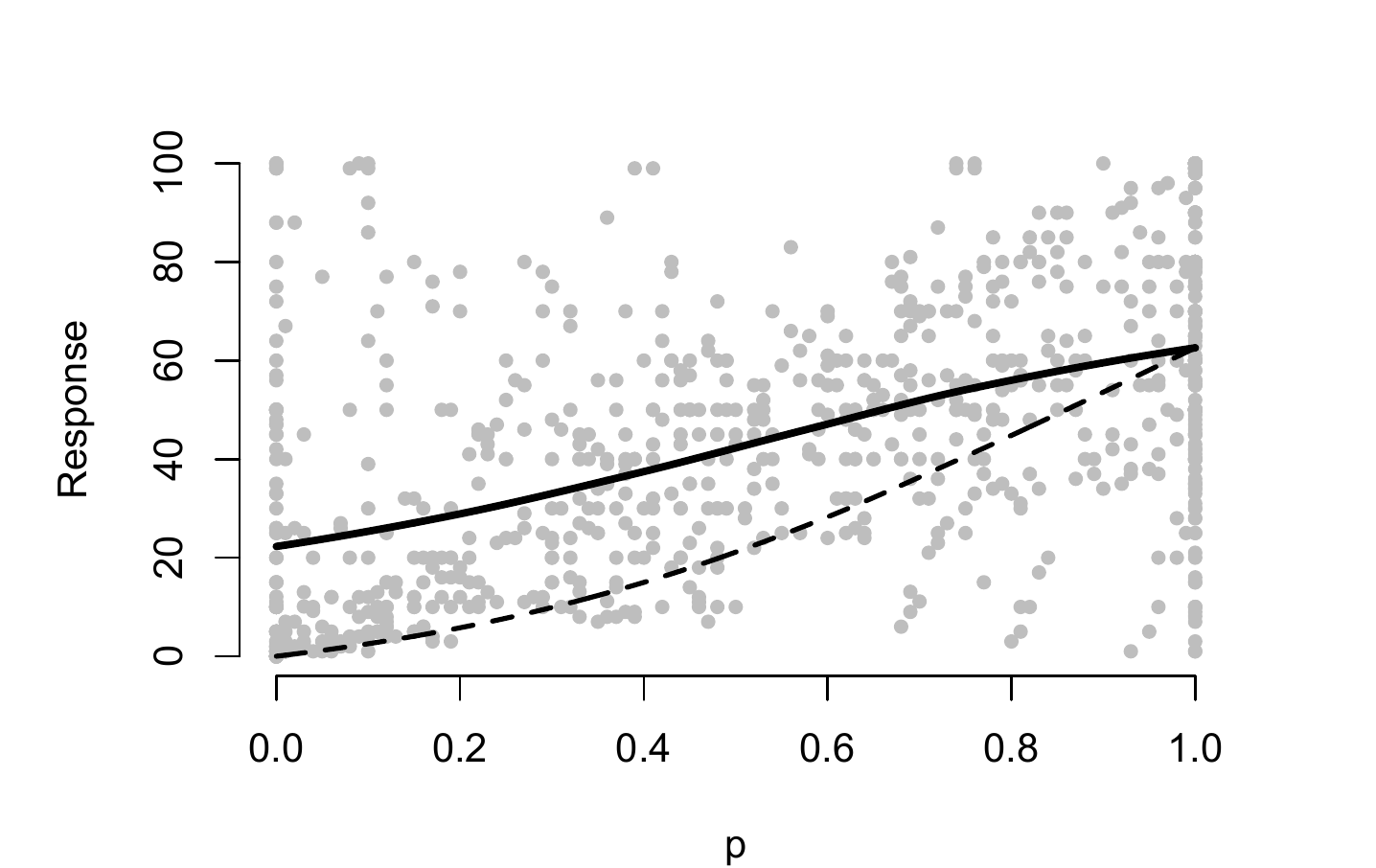}
\end{center}
\caption{Responses of 100 subjects over eight one-shot beauty contests. The smoothed empirical mean is shown in solid and the {\em ex post} optimal strategy is shown dashed. Note the non-Nash plays at $p = 0$.} \label{expost}
\end{figure}

A preliminary round of data collection was obtained where participants were randomly assigned values of $p \in [0,1]$ (including exactly 0).  This data is depicted in the scatter plot shown in Figure \ref{expost}.  This preliminary study was used to ascertain a reasonable class of regression functions to use for modeling the player strategies and to help determine  (informally) what fixed values of $p$ to use in the second phase of collection.  In the end, each subject played the game for $p \in \{0.3, 0.4, 0.5,0.6,0.7,1.0\}$.

All participants were paid a base fee of \$0.25 for participation, regardless of the game outcome.  The winner of each game (for each value of $p$) was awarded a \$50 bonus.  We also conducted a distance-based payoff version of the game, but, for clarity and brevity, we do not report those results here as the substantive conclusions were comparable.

Our subject pool was restricted to Amazon Turk users in the United States with a demonstrated track record of successful task completion\footnote{The pilot data collection was restricted to India.}.  See \cite{turk} for general technical details about Amazon Turk recruitment procedures.

Game instructions were provided in the form of a voice-over video with visual aids, which can be seen at {\tt http://faculty.chicagobooth.edu/richard.hahn/instructions.html}. Subjects were not explicitly told how many opponents they would be playing against, but were told they would be playing against other Amazon Turk workers\footnote{We presume that Turk participants are used to large numbers of other participants in this context.  On the other hand, the instructional video demonstrates the game with only four players.}.

Subjects were given an attention task at the end of the game instructions. Subjects were instructed to answer a post-game questionnaire item asking for their favorite color by entering the word ``SQUARE".  This allowed us to screen for participants who did not watch the instructional video to its completion.

\begin{figure}[h!]
\begin{center}
\includegraphics[width=6.5in]{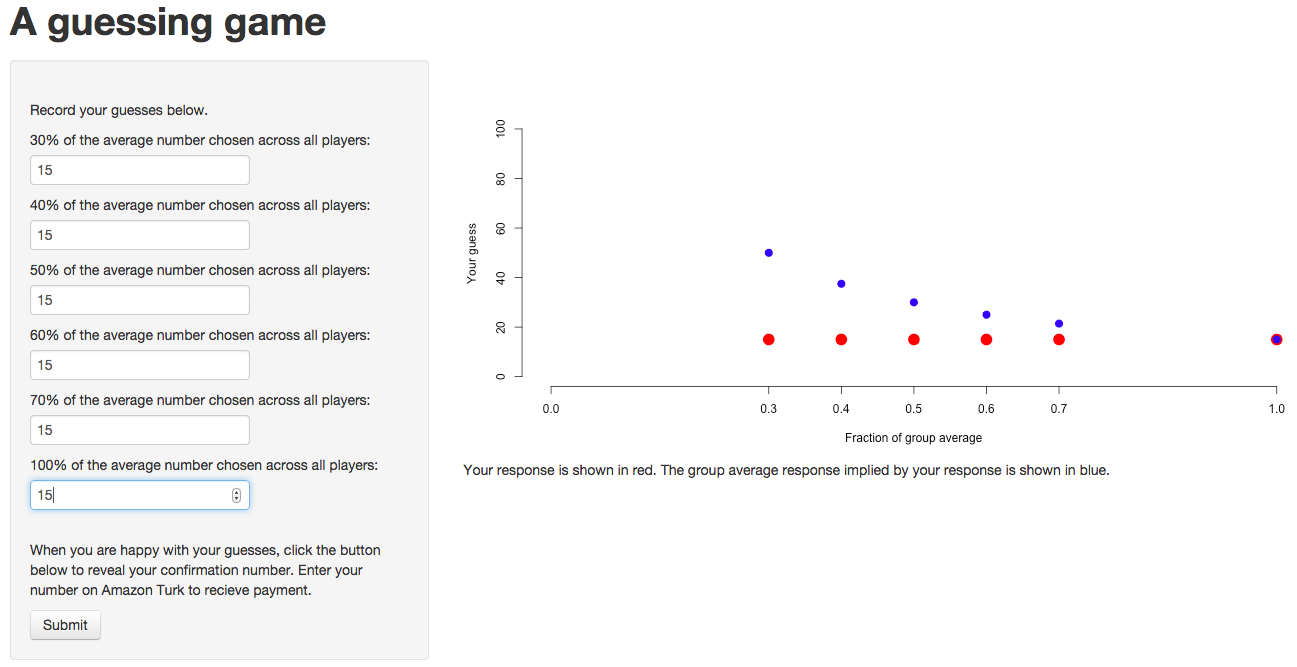}
\end{center}
\caption{Game interface as presented to study subjects online. The form is presented to subjects empty to start.  Here, to demonstrate the plotting functionality, it is shown configured as if a player responded with the number 15 for every value of $p$.} \label{screenshot}
\end{figure}

During the game, players were provided visual feedback of their picks across different values of $p$ via a web interface.   A screen shot of this interface can be seen in Figure \ref{screenshot}.  The interface allowed adjustment of all six numbers (one for each value of $p$), in any order, prior to submission. The web form automatically recorded the total time taken to complete the games. Simultaneous submission, along with the graphical interface, permitted users to adjust their game play to be mutually coherent without having to remember their responses for previous values of $p$.  This feature was introduced to ease concerns about the influence of an ordering effect \citep{hogarth1992order}, whereby the order in which players are presented with the various values of $p$ could systematically affect players' approach to the game.   

The post-game questionnaire collected additional demographics on each subject, including age, gender, and level of education.  Also included was a question adapted from the Cognitive Reflection Task \citep{Frederick2005}, consisting of the following elementary (but non-trivial) algebra problem:

\begin{quote}
A bat and ball cost \$30.10 in total.  The bat costs \$25 more than the ball.  How much does the ball cost?
\end{quote}
The time-taken to complete the questionnaire was also recorded.

\subsection{Data summaries}
In this section, we present some summary statistics from our data collection effort.  These summaries serve to characterize our study population and thus will help put our findings in context.  We also show example game play data, to illustrate the variety of patterns observed.

\begin{figure}
\begin{center}
\includegraphics[width=3.5in]{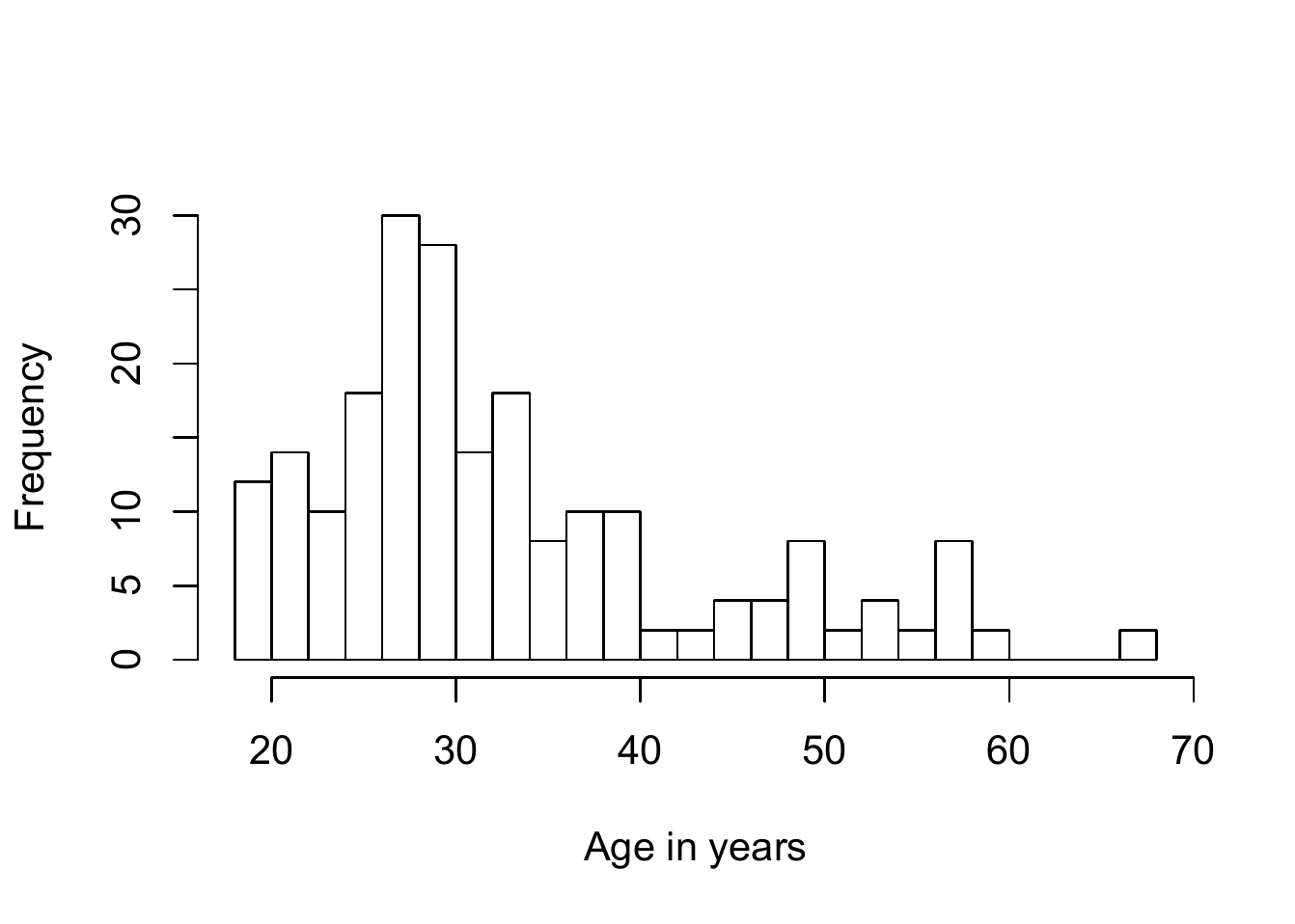}\includegraphics[width=3.5in]{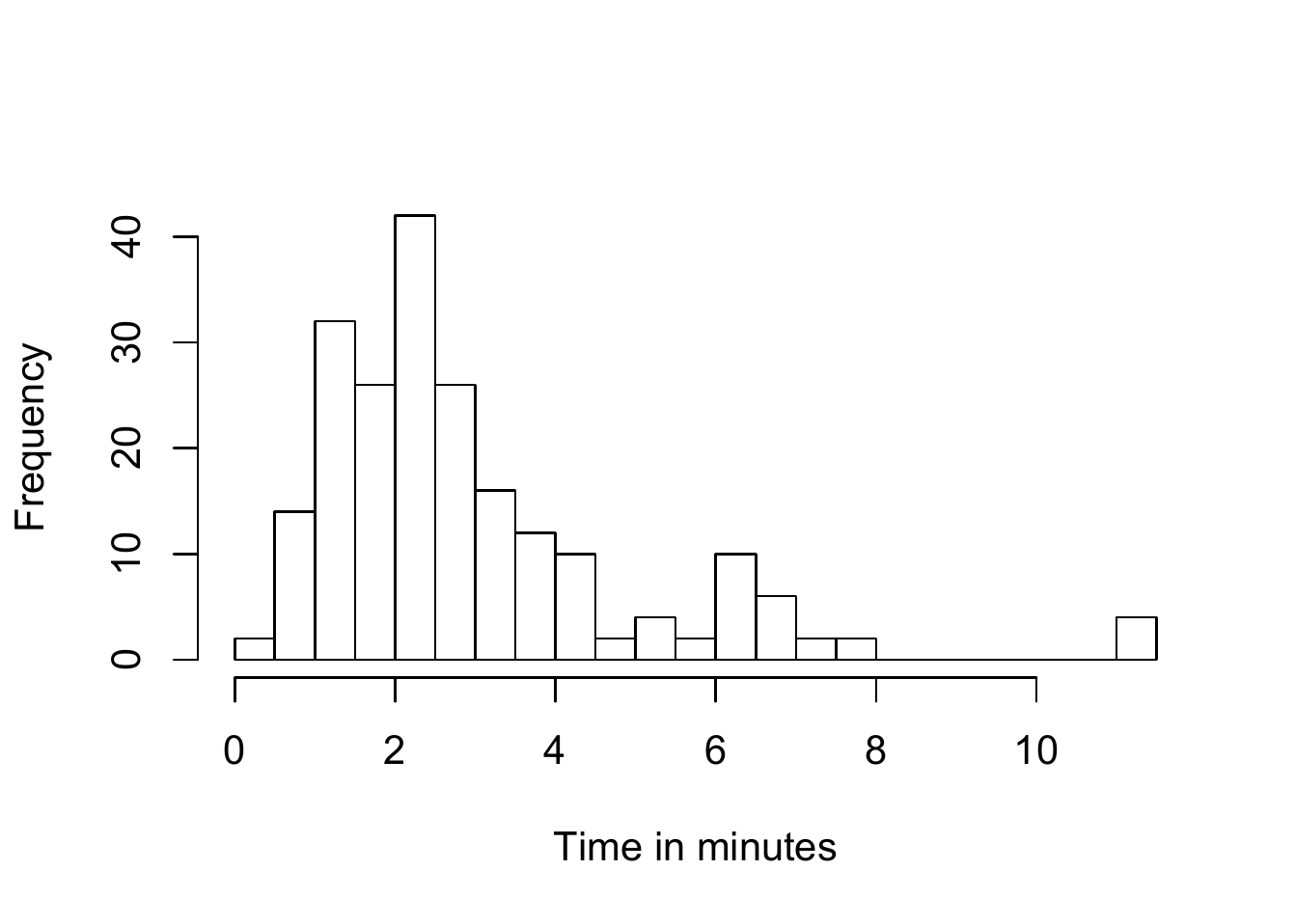}\\
\includegraphics[width=3.5in]{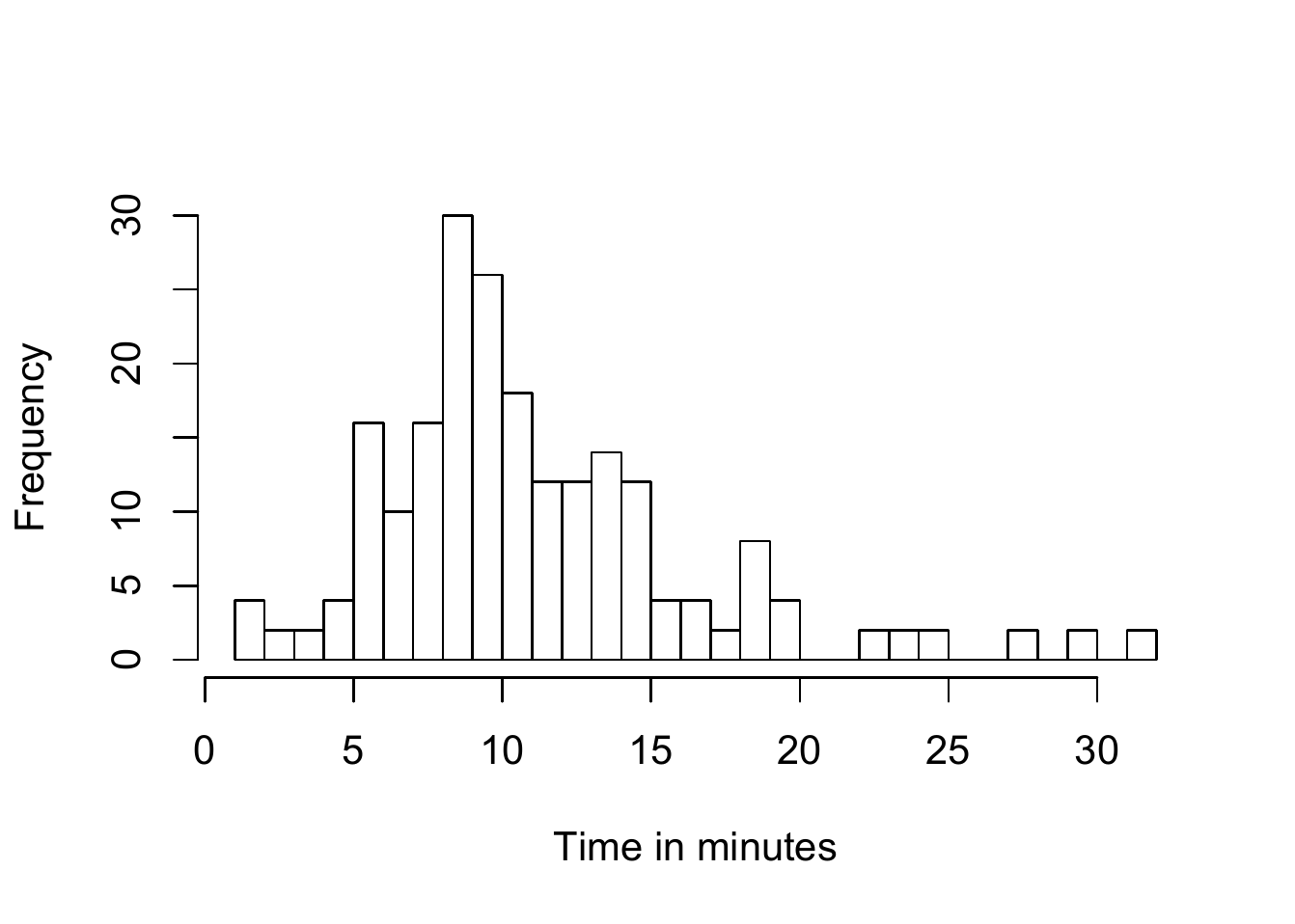}
\end{center}
\caption{Histograms of (from left to right) subject age, time to complete all six games,  and the time to complete both the six games and the questionnaire.}\label{hists}
\end{figure}

In the final collection, we had $n = 106$ subjects, 56\% female and 44\% male, with median age of 30. Education level statistics are given in Table \ref{edulevel}. While 80\% satisfied the attention task, only 34\% passed the cognitive reflection task. Figure \ref{hists} shows histograms of age, time to complete all six games, and combined time to complete games and questionnaire.  It is possible that our subject pool is unrepresentative---in one sense or another---of the prototypical strategic reasoner.  To be conservative, our inferences should be interpreted as applying narrowly to the population of U.S.-based Amazon Turkers. We leave unaddressed the question of how best to extrapolate our findings to other populations, but maintain that extending our analysis to other populations would be straightforward with additional data.

\begin{table}
\caption{Educational statistics of our study population.}\label{edulevel}
\begin{center}
\begin{tabular}{lr}
Education & Prevalence (\%)\\
\hline
Associates deg. & 8.5\\
Bachelors deg. & 36.0\\
Graduate deg. & 13.0\\
High school diploma &7.5\\
Some college &31.0\\
Some high school& 4.0
\end{tabular}
\end{center}
\end{table}

\begin{figure}
\begin{center}
\includegraphics[width=6in]{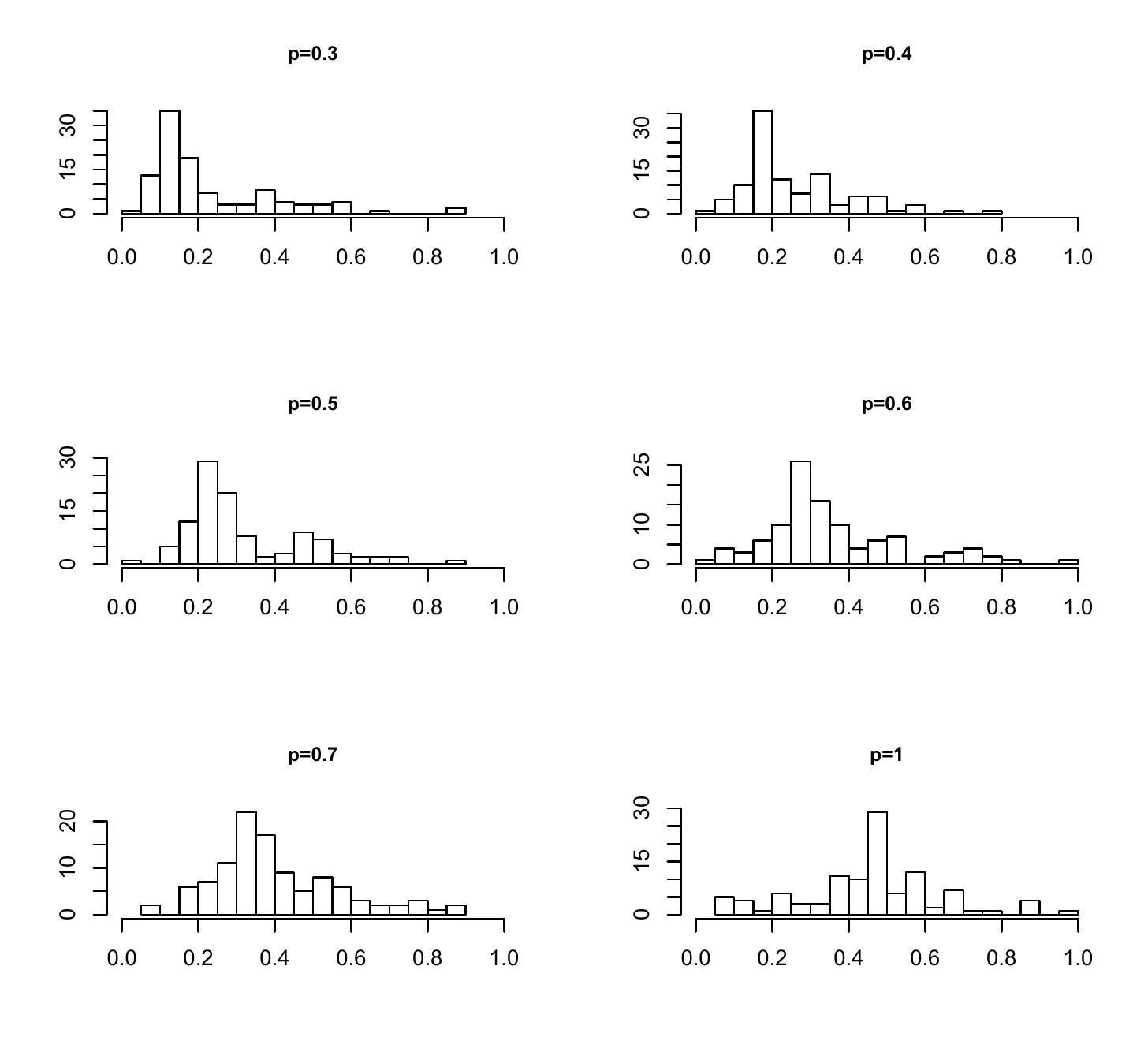}
\end{center}
\caption{Histograms of player responses for each value of $p$. Responses are rescaled to the unit interval.}\label{hist_raw}
\end{figure}

Figure \ref{hist_raw} shows histograms of player responses at each value of $p$.  This has been a standard way to visualize beauty contest data.  Observe that the modes in the data appear to migrate upwards with higher values of $p$.  Notice also a clear mode near $1/2$ in the $p=1$ data.

\begin{figure}
\begin{center}
\includegraphics[width=3.2in]{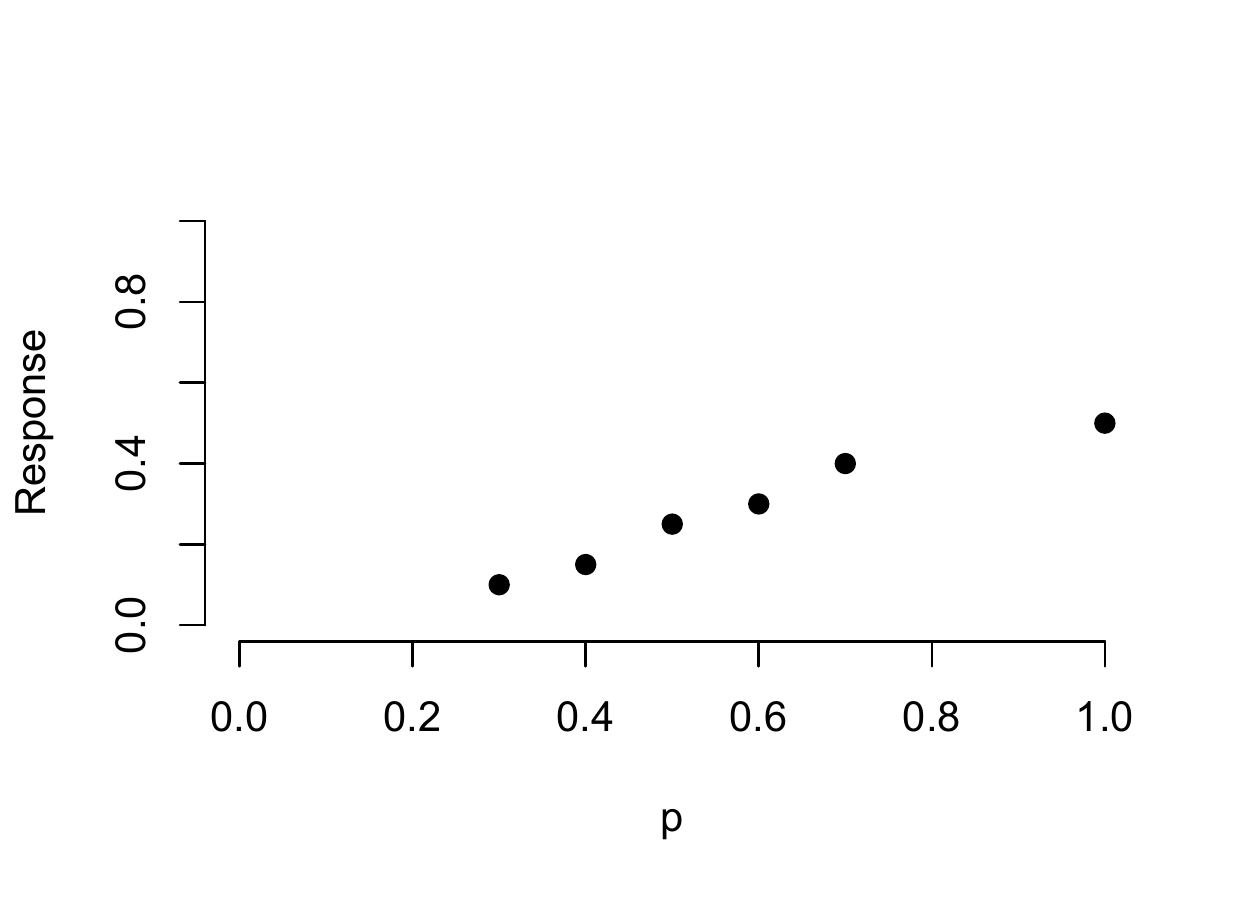}
\includegraphics[width=3.2in]{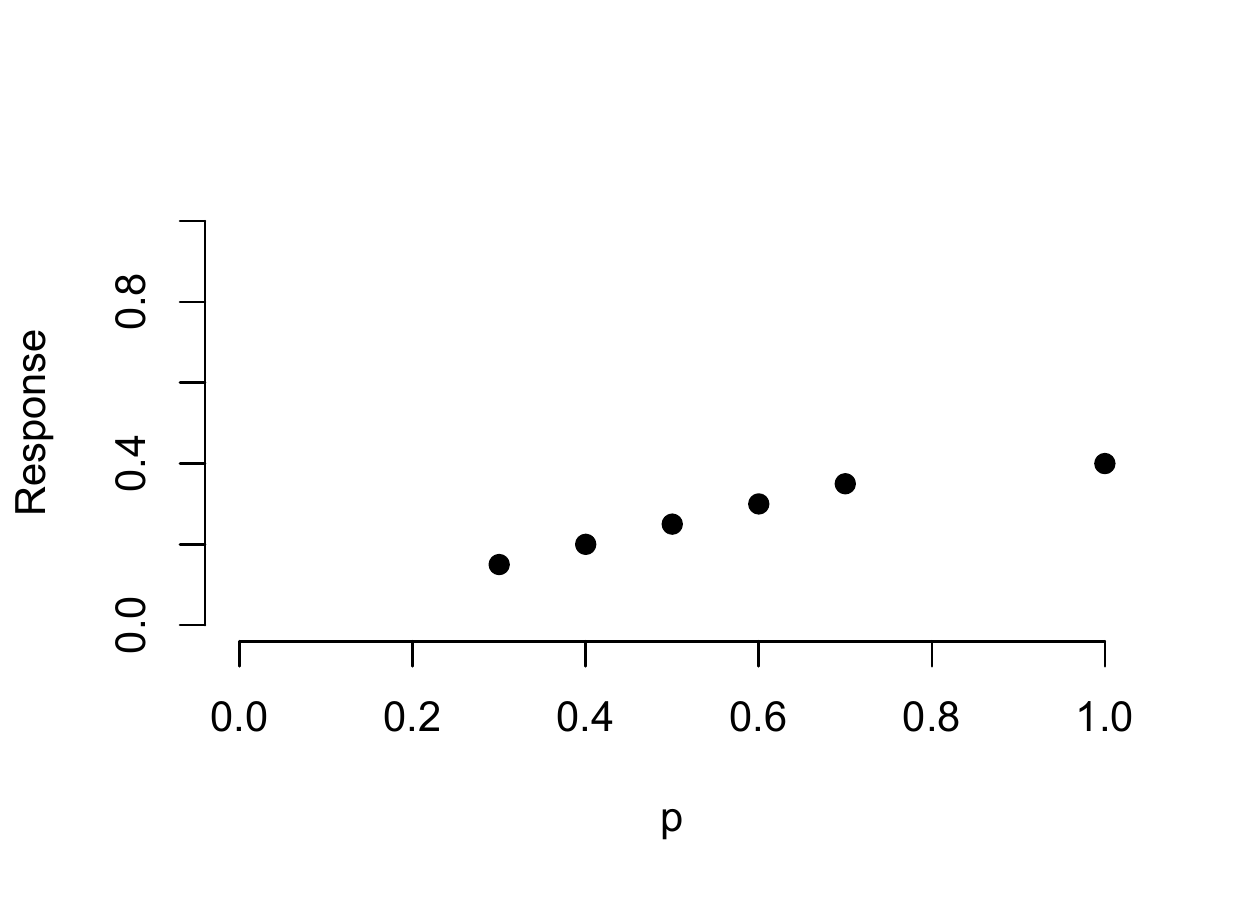}
\includegraphics[width=3.2in]{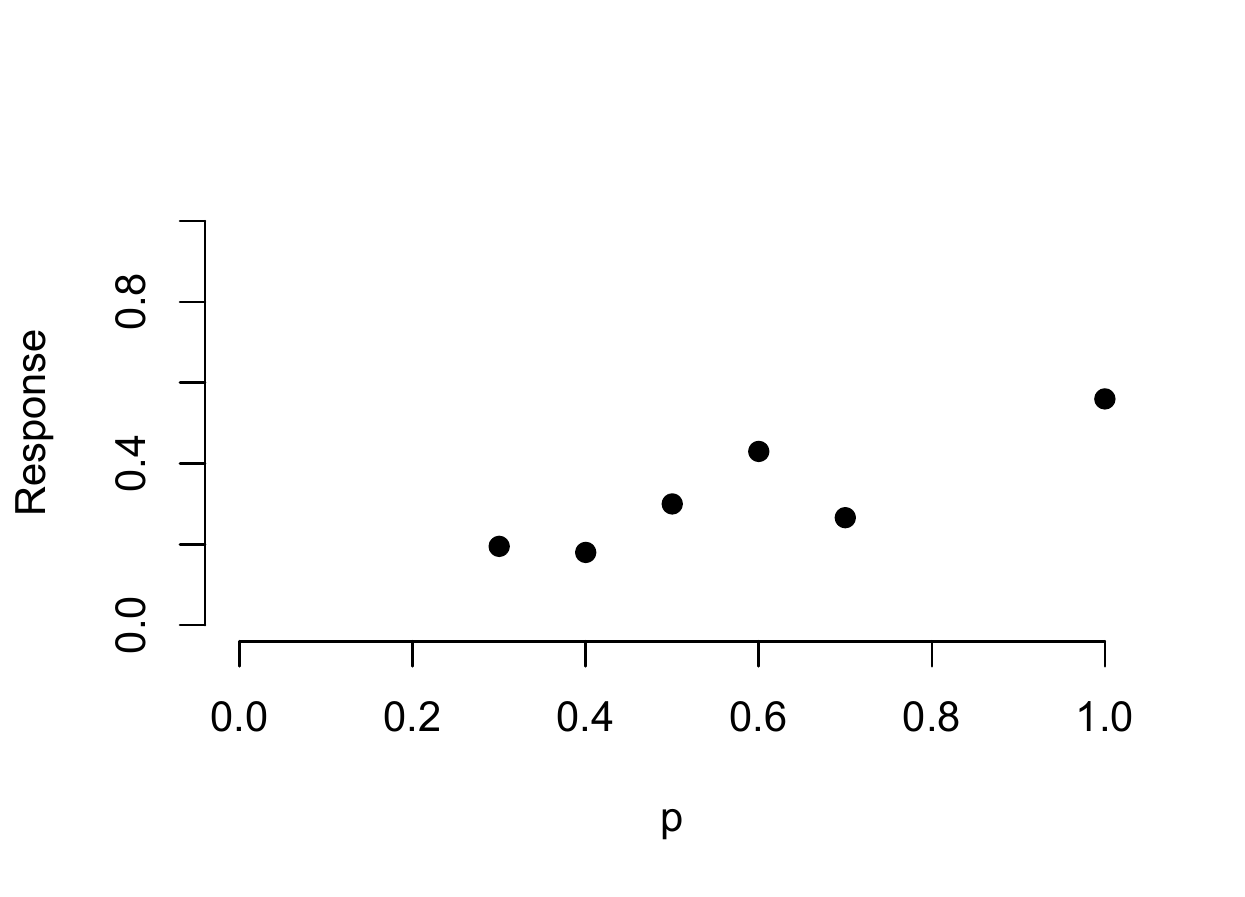}
\includegraphics[width=3.2in]{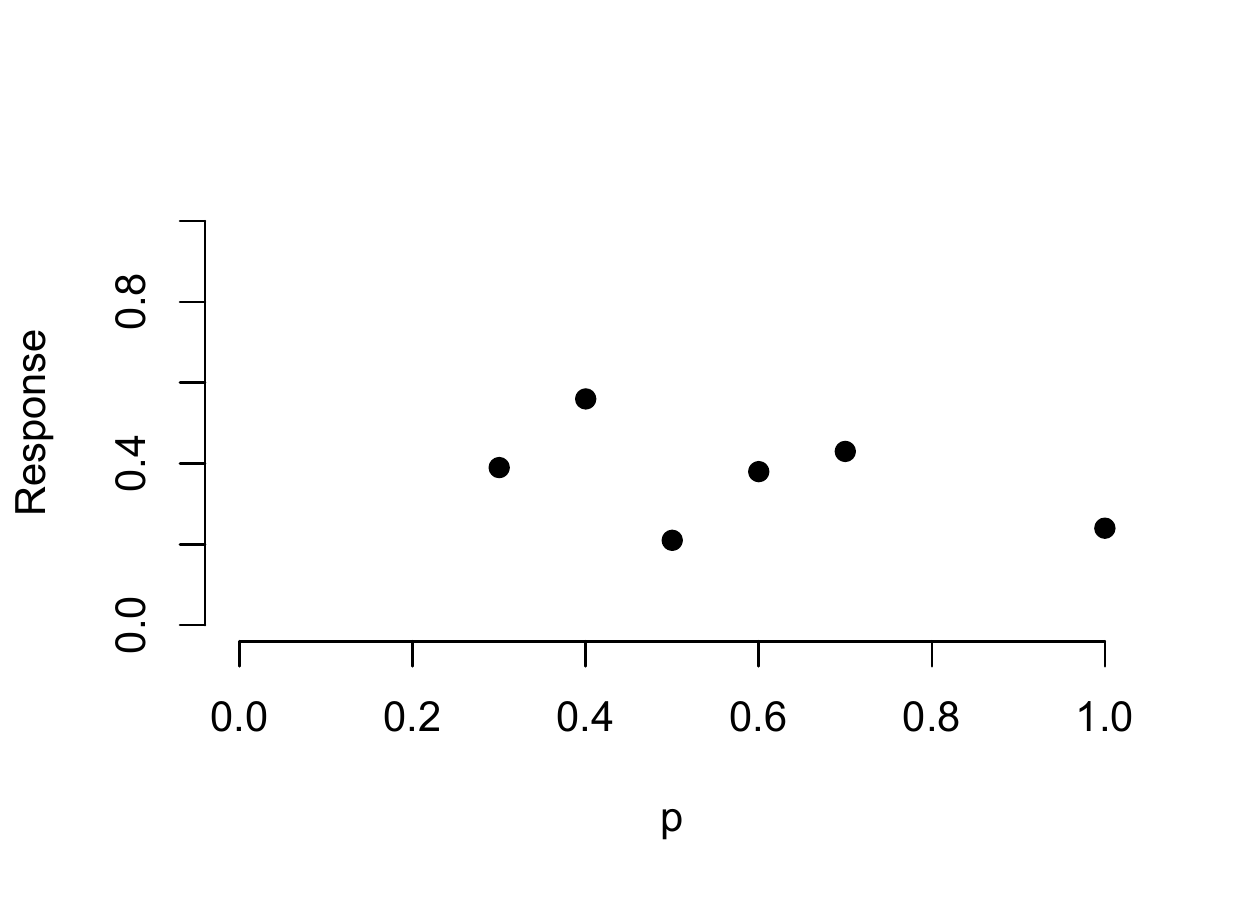}
\end{center}
\caption{Game data from four subjects. Responses are rescaled to the unit interval.}\label{data_ex}
\end{figure}

For data with multiple games per subject without feedback, such as ours, it is perhaps more illuminating to look at the data in a scatterplot, per subject.  Figure \ref{data_ex} shows four sets of data collected from representative subjects.  The fact that responses typically increase with $p$  suggests that subjects are generally attending to the experimental task.

\section{A statistical model for beauty contest data}\label{stats_model}
Our statistical analysis follows from a hierarchical model in which each individual has their own mean regression curve, which describes their strategy as a function of the game parameter $p$.  These individual strategies share a common prior distribution with unknown hyperparameters. 

In this section, we provide the details of this model, including prior specification\footnote{Computational details are given in Appendix A.  Trace plots and other diagnostic plots of key model parameters are provided in Appendix B.}.  We specify our model in terms of three components:  a model for random variation in players' responses, a model for each player's strategy as a function of $p$, and a description of the hyperparameters that are shared across players.

\subsection{Random variation in player responses}

We allow that players' responses can deviate from their underlying strategy owing to inattention or other unobserved contextual factors\footnote{We do not suppose that players entertain the possibility that their opponents will bid with error.  This possibility can be accommodated with additional assumptions on the nature of the errors, but we do not further consider this issue here.}.   Specifically, we assume that player responses $y_i(p)$ (rescaled to the unit interval) arise as draws from a Beta distribution.  Letting $\mu_i(p) \equiv \E\lbrace y_i(p)\rbrace$, our player response model is
\begin{equation}\label{likelihood}
y_i(p) \mid \mu_i, s_i \sim  \mbox{Beta}(c_i\mu_i(p),c_i(1-\mu_i(p))).
\end{equation}
To restrict this model to be unimodal, we define $c_i \equiv s_i \times \mbox{max}(1/\mu_i, 1/(1-\mu_i))$ with $s_i > 1$.  In this parametrization, $s_i$ controls the precision of player response distributions; higher values of $s_i$ correspond to higher values of $c_i$, which correspond in turn to density functions that are more sharply peaked about the mean, $\mu_i(p)$.  

\begin{figure}
\begin{center}
\includegraphics[width=5.5in]{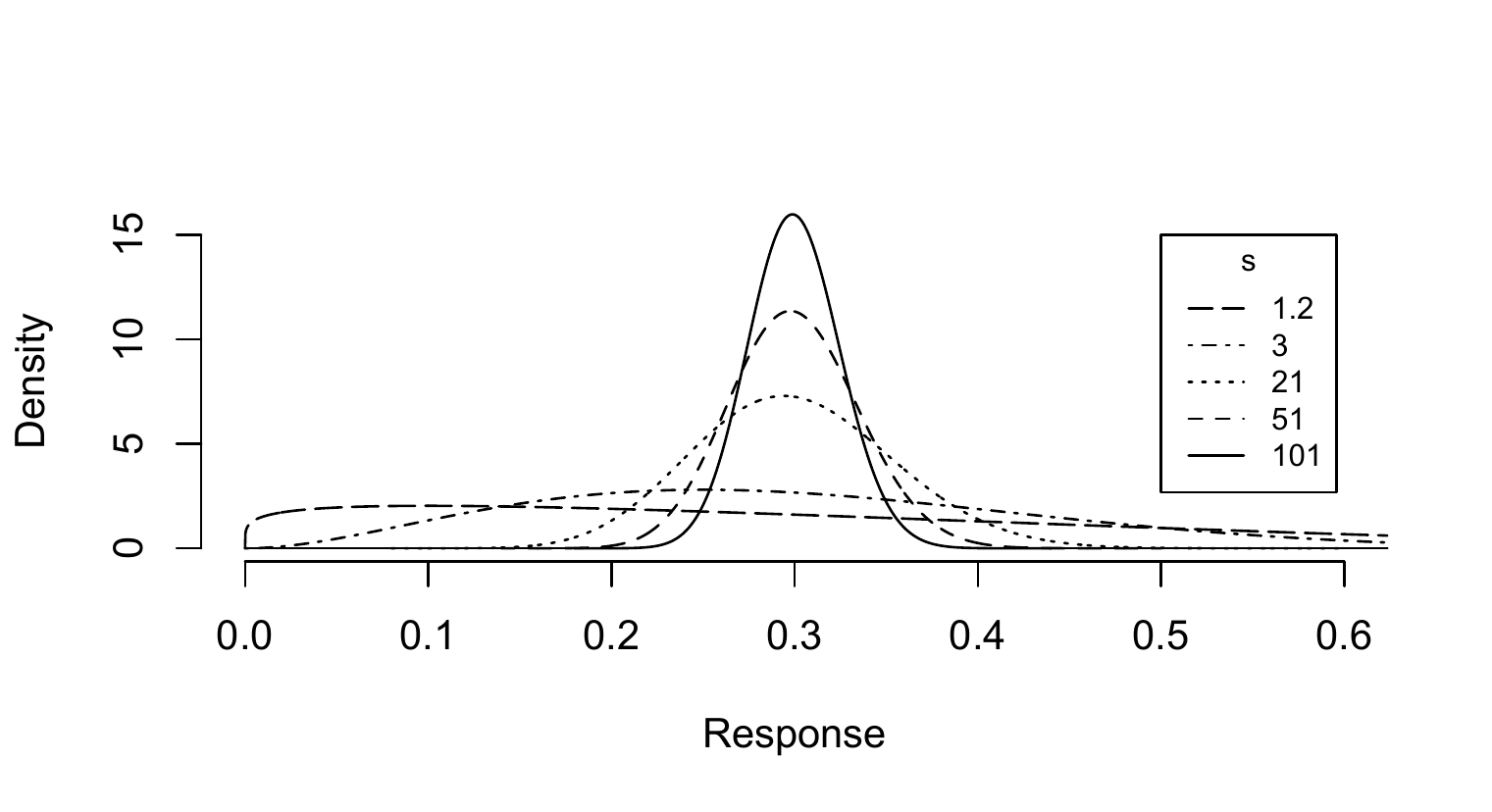}
\caption{Players can be one of five types, differing in the amount of precision with which they play about their mean strategy.  Here, the densities of the five types are shown for their corresponding values of $s$ when the mean strategy is fixed at 0.3.}\label{errors}
\end{center}
\end{figure}

We place a discrete prior on $s_i$, defined by the five possible values, $\lbrace 1.2, 3,21,51,101 \rbrace$, and a probability vector $\bf{w}$.  The choice of these values gives a wide range of player ``types" ranging from imprecise to precise, as shown in Figure \ref{errors}. The prior on $\bf{w}$ is discussed in Section \ref{hyperprior}.

\subsection{Mean response (strategy) curves}
The $k$-step compatibility criteria from Section \ref{estimation} can now be stated directly in terms of the expected player response function $\mu_i(p)$: i) $\mu_i(0)=0$, and ii) $\mu_i(p)$ is convex and increasing in $p$.  To facilitate verifying that these criteria are satisfied, we specify $\mu_i(p)$ in terms of a four-parameter vector, $\theta_i \equiv (\eta_i, \phi_i, \nu_i, \mu_i^0)$, using a spline representation. (In the following discussion, we suppress subscripts for convenience.)  

\begin{figure}
\begin{center}
\includegraphics[width=5.25in]{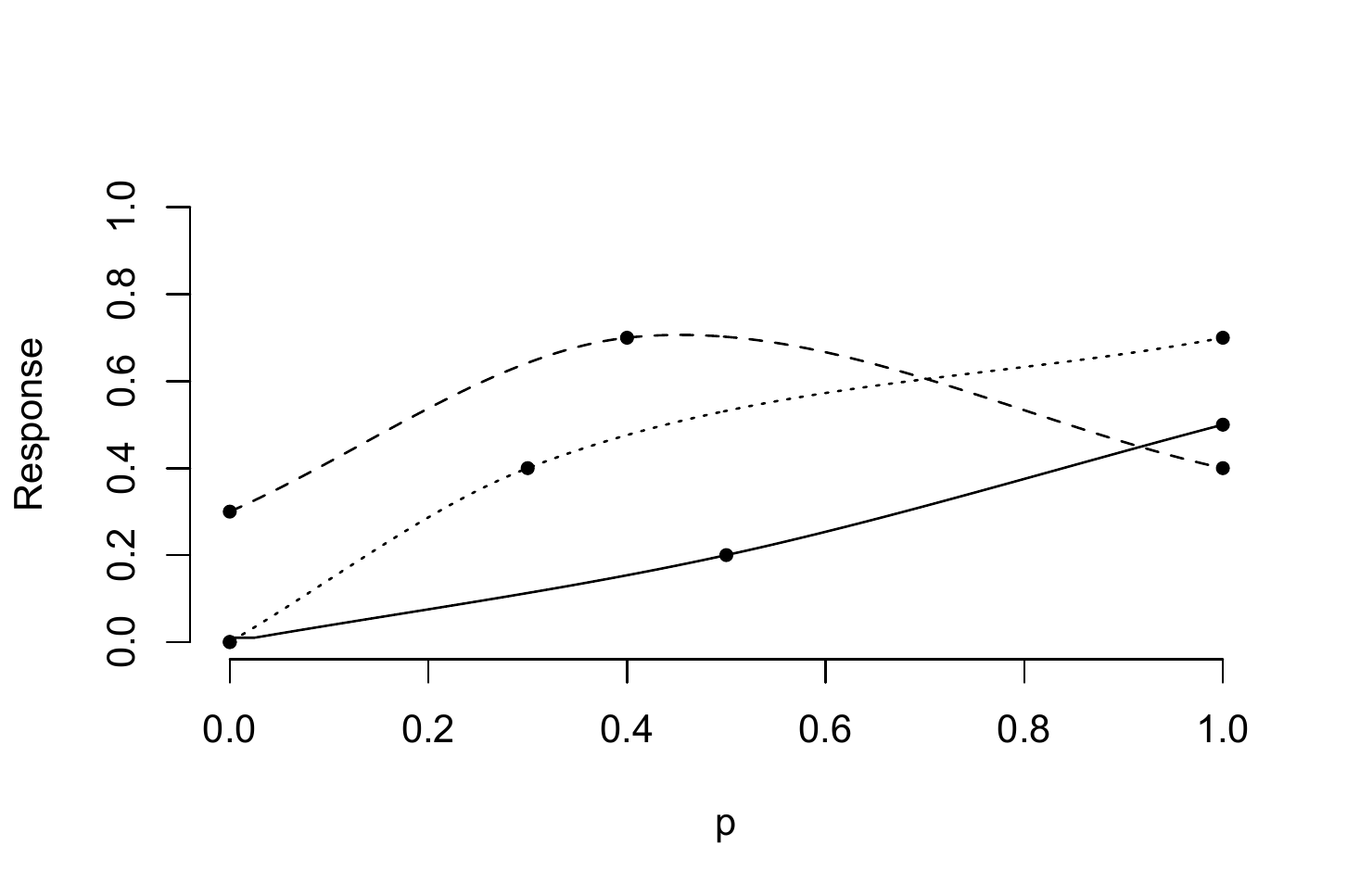}
\end{center}
\caption{Three curves are shown with associated parameters $\theta_{dashed} = (0.4,0.3,0.7,0.4), \theta_{dotted} = (0.5, 0,0.2,0.5)$ and $\theta_{solid} = (0.3,0,0.4,0.7)$ shown as points. The dashed curve neither intersects the origin nor is convex.  The dotted curve intersects the origin but is not convex.  The solid curve both intersects the origin and is convex.  Only the solid curve could represent a $k$-step thinking strategy.} \label{splines}
\end{figure}

Intuitively, $\theta$ defines three {\em control points} in the $(p, y)$ plane: $(0, \phi)$, $(\eta, \nu)$ and $(1, \mu^0)$.   The function $\mu(p)$ is determined by drawing a smooth monotone curve through these points.  By monotone\footnote{We implement the spline component of the model via the {\tt R} function {\tt splineFun()} with setting {\tt method = `mono'}, which is based on the method of \cite{fritsch1980monotone}.}, we mean that $\mu(p)$ is monotonically increasing if and only if $\phi \leq \nu \leq \mu^0$. The left control point, $(0, \phi)$, controls whether or not $\mu(p)$ intersects the origin, which happens precisely when $\phi = 0$.  Likewise, the right control point is given by $(1, \mu^0)$.  Now, imagine drawing a straight line between $(0, \phi)$ and $(1, \mu^0)$.  If the interior control point, $(\eta, \nu)$, lies below this line, then $\mu(p)$ is convex, otherwise it is not. In other words, $\mu_i(p)$ will be convex if and only if 
\begin{equation}\label{inequality}
\nu  <  (\mu^0 - \phi)\eta + \phi.
\end{equation}
This inequality allows us to control the prior probability of a convex strategy via the prior for $\nu_i$, conditional on the other three spline parameters ($\phi, \eta, \mu^0$). Example curves are shown in Figure \ref{splines}.  

\newpage
\subsubsection{Prior for $\mu_i(p)$}
The prior on each $\mu_i(p)$ is induced via priors on the parameters $\eta_i, \phi_i, \nu_i$, and $\mu^0_i$.\\

\noindent {\em Priors over $\mu_i^0$ and $\eta_i$}\\
First, $\mu_i^0$, which defines the right control point $(1, \mu^0)$, is given an independent $\mbox{Beta}(3/2,1)$ prior.  Similarly, $\eta_i$, which determines the horizontal location of the interior control point $(\eta_i, \nu_i)$, is given an independent $\mbox{Beta}(5,5)$ prior, rescaled to be supported on interval $[0.3, 0.7]$. This restricted support improves numerics by preventing linear trends from being fit by $\eta_i$ values near 0 or 1. \\

\noindent {\em Prior over $\phi_i$}\\
Next, $\phi_i$ defines the left control point $(0, \phi)$.  To control the probability that $\mu_i(p)$ runs through the origin, we place a prior over $\phi_i$ with a point mass at zero
\begin{equation}\label{phi_prior}
\phi_i \mid \rho, \mu^0_i \sim \rho \delta_0 + (1-\rho) \mbox{Beta}(c\mu^0_i,c(1-\mu^0_i))
\end{equation}
and $c = 2 \max{(1/\mu^0_i,1/(1-\mu^0_i))}$.  This is a zero-inflated mixture with weights $\rho$ and $1-\rho$.  We condition on $\mu_0$ so that the continuous portion of the prior specifies $\E(\phi \mid \mu_i^0, \phi \neq 0) = \mu_i^0$.  The parameter $\rho$ thus determines the probability that a randomly selected strategy runs through the origin.\\

\noindent {\em Prior over $\nu_i$}\\
Finally, we describe a prior over $\nu_i$, the vertical component of the interior control point $(\eta_i, \nu_i)$.  We specify this prior in terms of a binary latent variable, $\kappa_i$, that designates if $\mu_i(p)$ is convex or not.  Specifically, if $\kappa_i = 1$, we must have that $\nu_i <  (\mu^0_i - \phi_i )\eta_i + \phi_i$; the interior control point lies below the line segment running between the left control point $(0, \phi)$ and the right control point $(1, \mu^0)$.  As we wish to remain uninformative about the likely values of $\nu_i$, we give it a uniform prior on $[0, (\mu^0_i - \phi_i )\eta_i + \phi_i]$.  By the same reasoning, if $\kappa_i = 0$, we give $\nu_i$ a uniform prior on $[(\mu^0_i - \phi_i )\eta_i + \phi_i, 1]$.

To complete this prior, we give the latent indicator of convexity, $\kappa_i$, a Bernoulli distribution that depends explicitly on the left control point $(0, \phi_i)$; strategy curves that intersect the origin can have different probabilities of being convex than strategies that do not intersect the origin.  That is, $\mbox{Pr}(\kappa_i = 1 \mid \phi_i = 0) \equiv q_0$ and $\mbox{Pr}(\kappa_i = 1 \mid \phi_i \neq 0) \equiv q_1$. \\

The above parametrization allows us to conveniently express the probability that a randomly selected strategy could have arisen from $k$-step reasoning: 
\begin{equation}\label{rho_q}
\mbox{Pr}(\kappa_i = 1, \phi_i = 0) = \mbox{Pr}(\phi_i = 0)\mbox{Pr}(\kappa_i = 1 \mid \phi_i = 0) = \rho q_0.
\end{equation}
Posterior inferences concerning prevalence of $k$-step thinking will thus follow directly from the posterior over these parameters.

\subsubsection{Priors over shared hyperparameters}\label{hyperprior}
\begin{figure}[h]
\begin{center}
\includegraphics[width=3.5in]{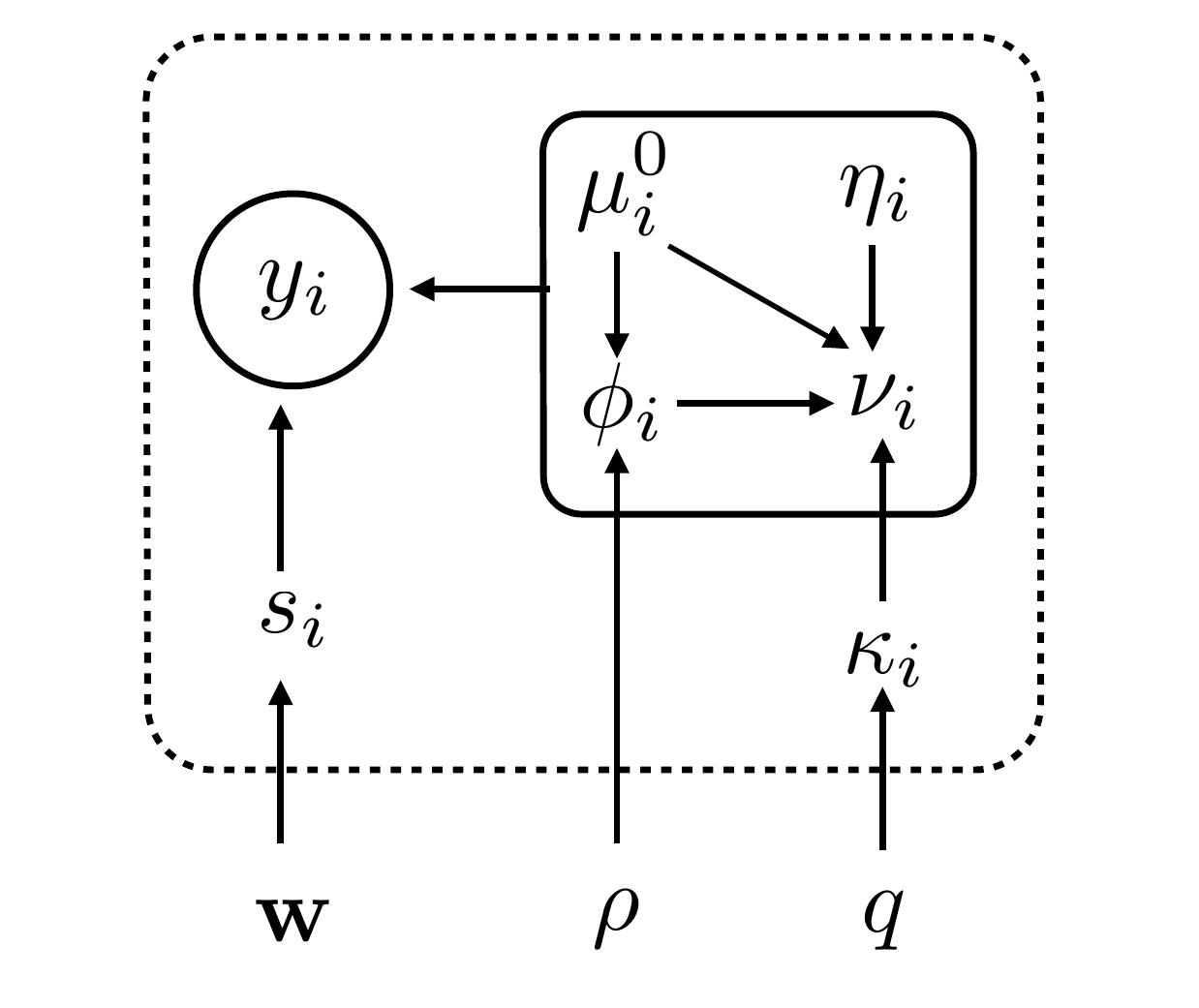}
\end{center}
\caption{A diagram of the dependence structure in our statistical model of beauty contest strategies curves and observed responses.  The dotted box indicates $n$ replicates, indexed by $i$.  The circled variable $y_i$ denotes the vector of observed responses. The solid box with rounded edges contains the parameters defining the mean response function $\mu_i(p)$ by monotone spline interpolation. The parameter $s_i$ controls the response error scale.  The hyperparameters $\mathbf{w}$, $\rho$, and $q = (q_0, q_1)$  are shared across all players.} \label{dag}
\end{figure}

Figure \ref{dag} illustrates the dependence structure between the data $y_i$, the parameters governing $\mu_i(p)$, $\lbrace \phi_i, \eta_i, \nu_i, \mu_i^0\rbrace$, and the shared hyperparameters, $\lbrace {\bf w}, q_0, q_1, \rho \rbrace$.  Although each player is permitted to have her own strategy, the shared hyperparameters shrink the individual estimates towards the group mean, mitigating the risk of over-fitting.  

Expression (\ref{rho_q}) shows that $\rho$ and $q_0$ directly quantify the prevalence of $k$-step compatible strategies. Hence, priors over these parameters have the potential to strongly influence posterior inferences.  For the analysis described in Section \ref{results}, we used $\rho \sim \mbox{Beta}(3,1)$, $q_0 \sim \mbox{Beta}(3,1)$ and $q_1 \sim \mbox{Beta}(1,3)$.  These choices reflect a favorable bias toward $k$-step compatible strategies, in that $\E(\rho q_0) = 9/16$, but are sufficiently diffuse to allow the data to contradict this bias.  %Figure \ref{posterior} depicts posterior learning of the key quantity $\rho q_0$, and shows clear movement from prior to posterior.

\begin{figure}
\begin{center}
\includegraphics[width=4.25in]{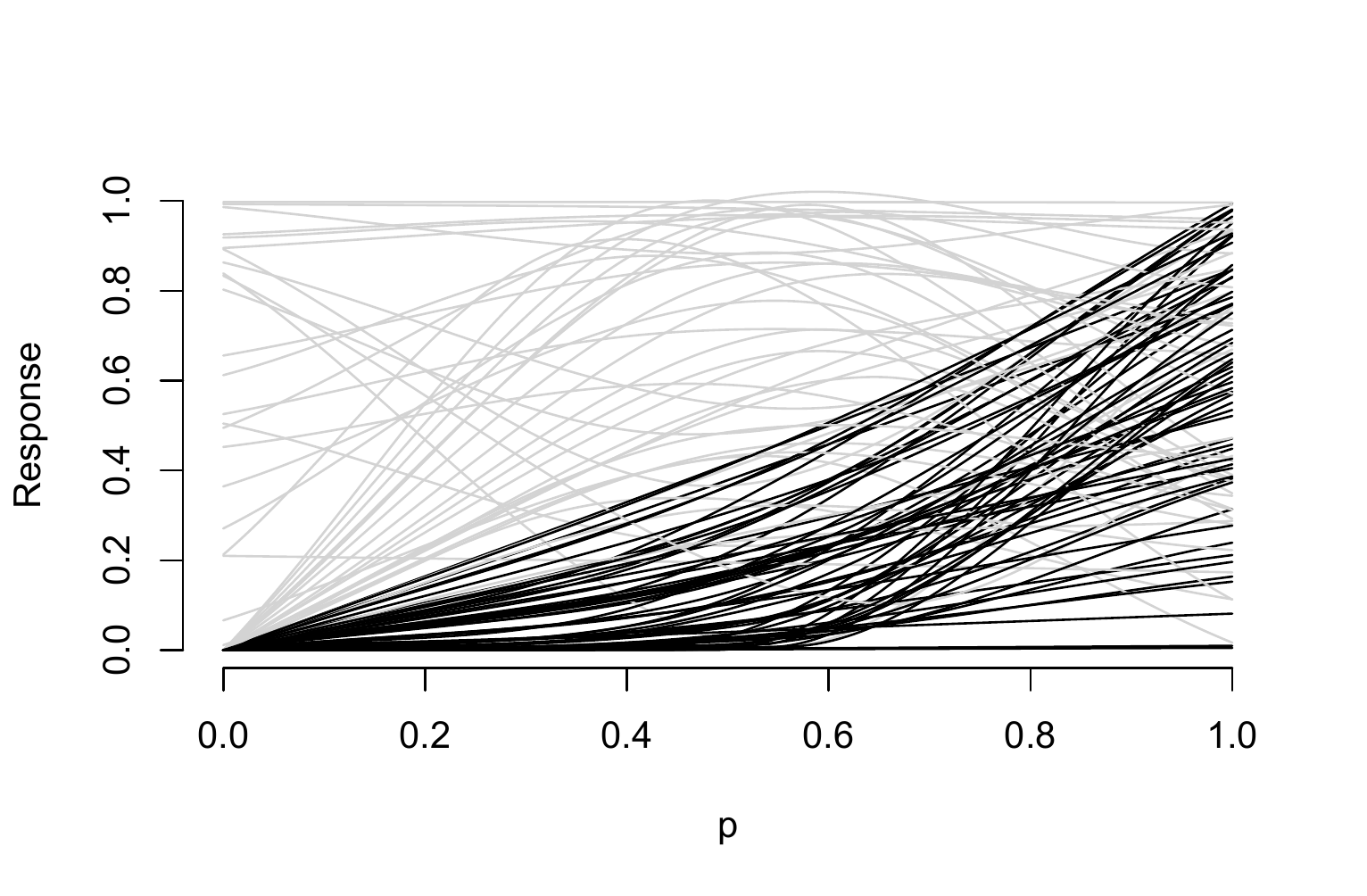}
\includegraphics[width=4.25in]{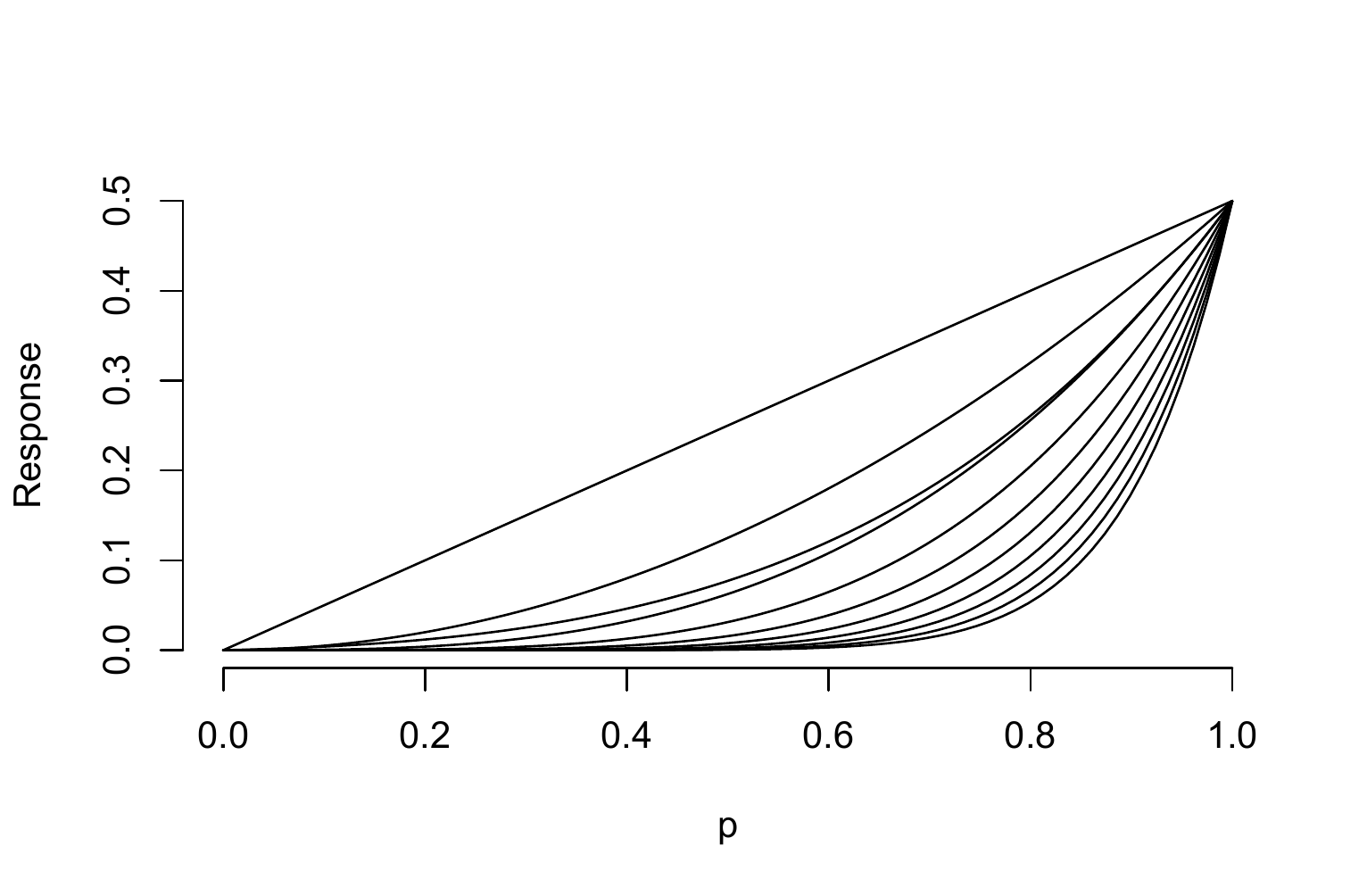}
\caption{The first panel shows one hundred randomly drawn strategies from the prior; $k$-step-compliant strategies are shown in black and non-$k$-step compatible strategies are shown in gray. For comparison, the second panel shows level-$k$ strategies (out to ten levels).}\label{prior}
\end{center}
\end{figure}

The above choices, as well as the priors over $\mu_i^0$ and $\eta_i$ described above, were guided by visual inspection of strategy curves drawn from the prior.  The first panel of Figure \ref{prior} illustrates many realizations from the prior over $\mu(p)$. Strategies consistent with $k$-step reasoning are shown in black, and non-$k$-step-compliant strategies are shown in gray.  

Lastly, the parameter $\bf{w}$ governs the prevalence of ``imprecise" players versus ``precise'' players in that it encodes the relative frequencies of more diffuse Beta distributions versus less diffuse (more concentrated) Beta distributions. We give $\bf{w}$ a Dirichlet prior with mode parameter $\boldsymbol{\alpha} = (0.1,0.25,0.3,0.25,0.1)$, which is symmetric and unimodal over the range of possible values, $\lbrace 1.2, 3,21,51,101 \rbrace$.  See again Figure \ref{errors} for a visualization of the Beta distributions implied by these values.

\section{Results and Future Work}\label{results}
The analysis below is based on 10,000 samples drawn from a Markov chain Monte Carlo simulation of our posterior distribution, after a burn-in period of 5,000 samples.  For details on our computational implementation and convergence diagnostics, see the Appendices.  The code to reproduce our plots, as well as the data to run it on, can be found online at {\tt http://faculty.chicagobooth.edu/richard.hahn/}.

\subsection{The prevalence of $k$-step compatible strategies}
\begin{figure}[htbp]
\begin{center}
\includegraphics[width=5.5in]{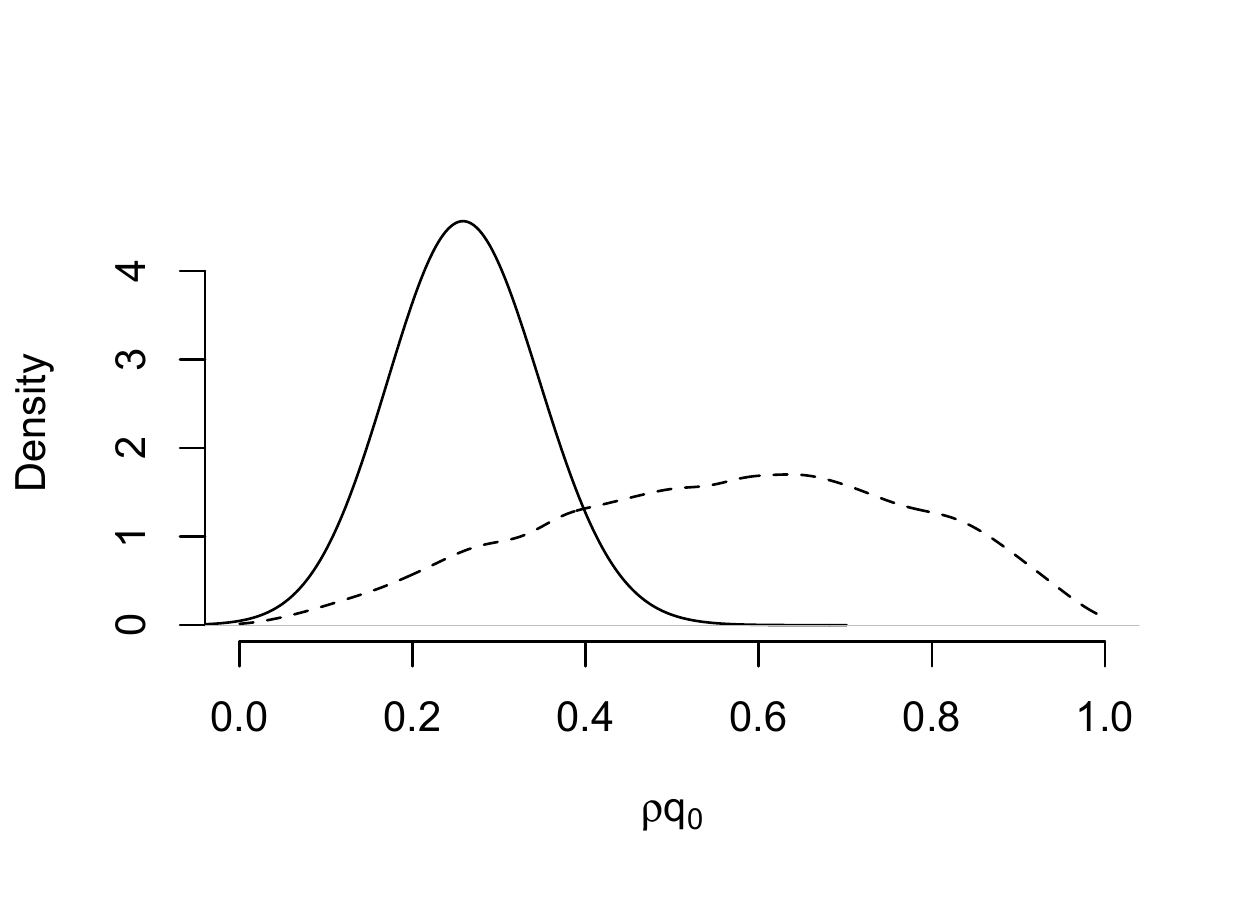}
\caption{The probability that a randomly selected participant plays a strategy that is $k$-step compatible is calculated as $\rho q_0$.  The solid line shows the posterior density of $\rho q_0$, and the dashed line shows the corresponding prior density.  The posterior mean is 26\%. }\label{posterior}
\label{default}
\end{center}
\end{figure}

Our main finding is that approximately 25\% of subjects appear to play $k$-step compatible strategies. Recall that $\rho q_0$ represents (an upper bound on) the probability that a randomly selected subject applies $k$-step reasoning.  The posterior mean of $\rho q_0$ is 26\%, with a 95\%-tile of 35\%.  By contrast, the prior mean was 56\% with a 95\%-tile of 89\%.  Figure \ref{posterior} depicts a kernel-smoothed Monte Carlo estimate of the posterior density of $\rho q_0$.  

\begin{figure}[htbp]
\begin{center}
\includegraphics[width=5.5in]{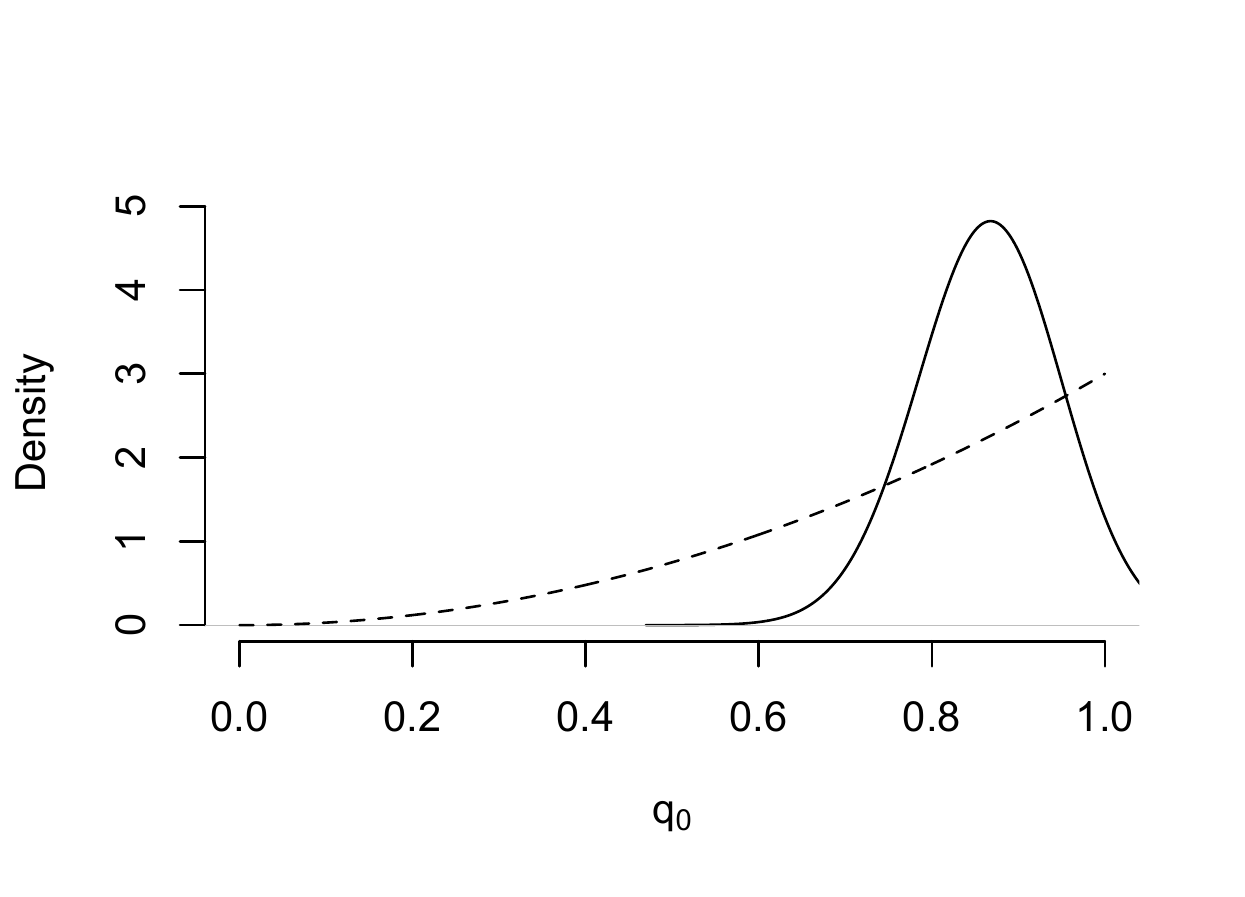}
\caption{The probability that a randomly selected participant plays a strategy that runs through the origin is given by the parameter $\rho$.  The solid line shows the posterior density of $\rho$, and the dashed line shows the corresponding prior density. The posterior mean is 87\%.}\label{rhoplot}
\label{default}
\end{center}
\end{figure}

On the other hand, a majority of player strategies seem to satisfy the first $k$-step compatibility criterion: intersecting the origin. Recall that $\rho$ represents the probability that a randomly selected participant's strategy intersects the origin. The posterior mean of $\rho$ is 87\%.  Figure \ref{rhoplot} depicts a kernel-smoothed Monte Carlo estimate of the posterior density of $\rho$.  

\begin{figure}[htbp]
\begin{center}
\includegraphics[width=5.5in]{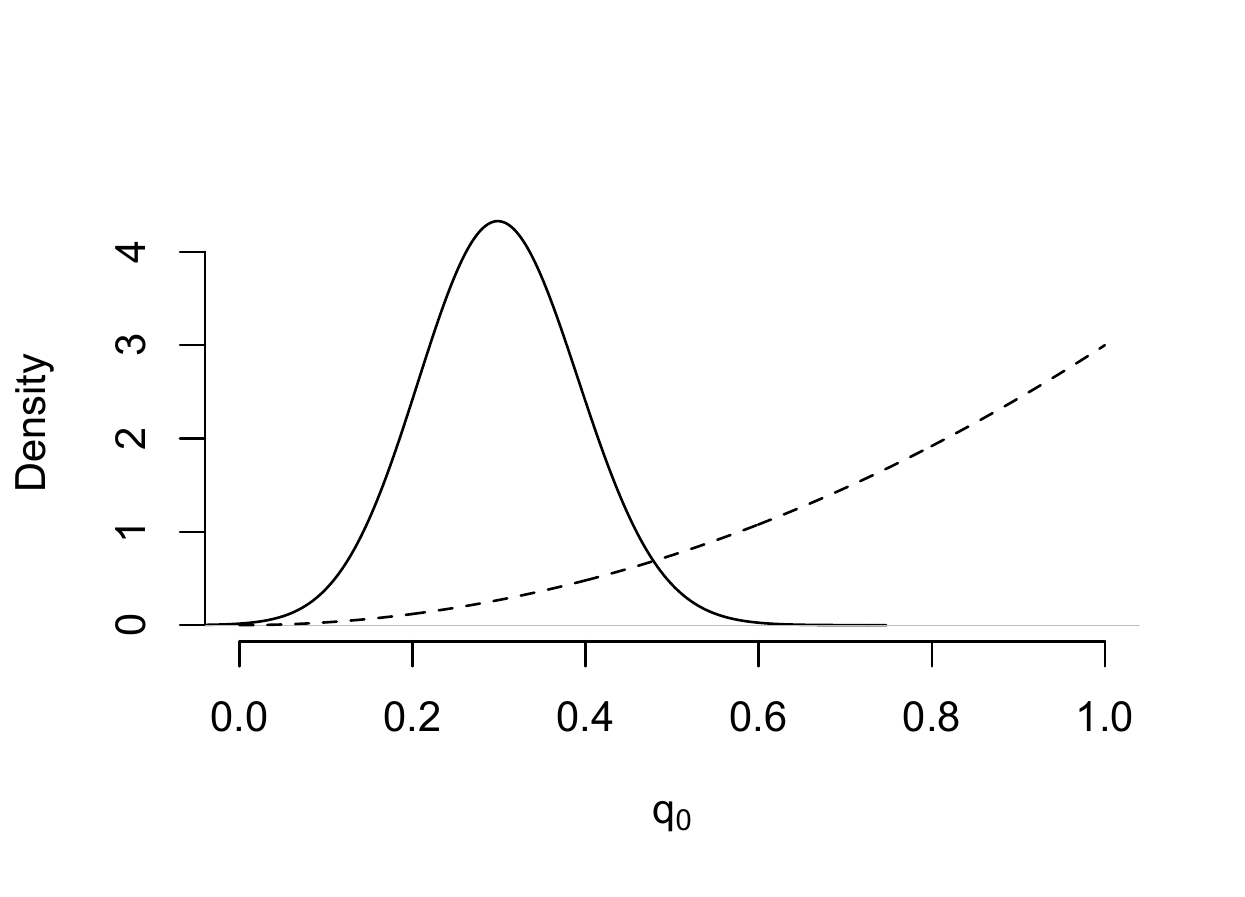}
\caption{The probability that a randomly selected participant plays a convex strategy, given that they play a strategy that intersects the origin, is given by the parameter $q_0$. The solid line shows the posterior density of $q_0$, and the dashed line shows the corresponding prior density. The posterior mean is 30\%.}
\label{posterior2}
\end{center}
\end{figure}

However, even among players whose strategy intersects the origin, convex strategies are not the norm.  Recall that $q_0$ represents the probability that a randomly selected participant plays a convex strategy, given that her strategy intersects the origin. The posterior mean of $q_0$ is 30\%, with a 95\%-tile of 40\%.   Figure \ref{posterior2} depicts a kernel-smoothed Monte Carlo estimate of the posterior density of $q_0$.  

Therefore, while some individuals may arrive at their strategies via $k$-step reasoning, many more appear to play strategies that do not correspond to any $k$-step derived strategy.  This finding is in broad agreement with that of \cite{coricelli2009neural}, who classify 7 out of 20 of their participants as greater than 1-step thinkers (35\%). Their classification is based on a simple least-squares based classification with no pooling of data.  To quantify what we mean by ``broad agreement", we compute the expected probability of drawing 7 $k$-step compatible strategies from a group of 20, if $\rho$ and $q_0$ were known; then, we take the average of this probability with respect to our posterior distribution over $\rho$ and $q_0$.  We find that, according to our posterior, observing the proportion of $k$-step compliant players reported in \cite{coricelli2009neural} is an expected probability of 12\%.
 
 \subsection{Individual-level posterior analysis}
\begin{figure}
\begin{center}
\includegraphics[width=3.2in]{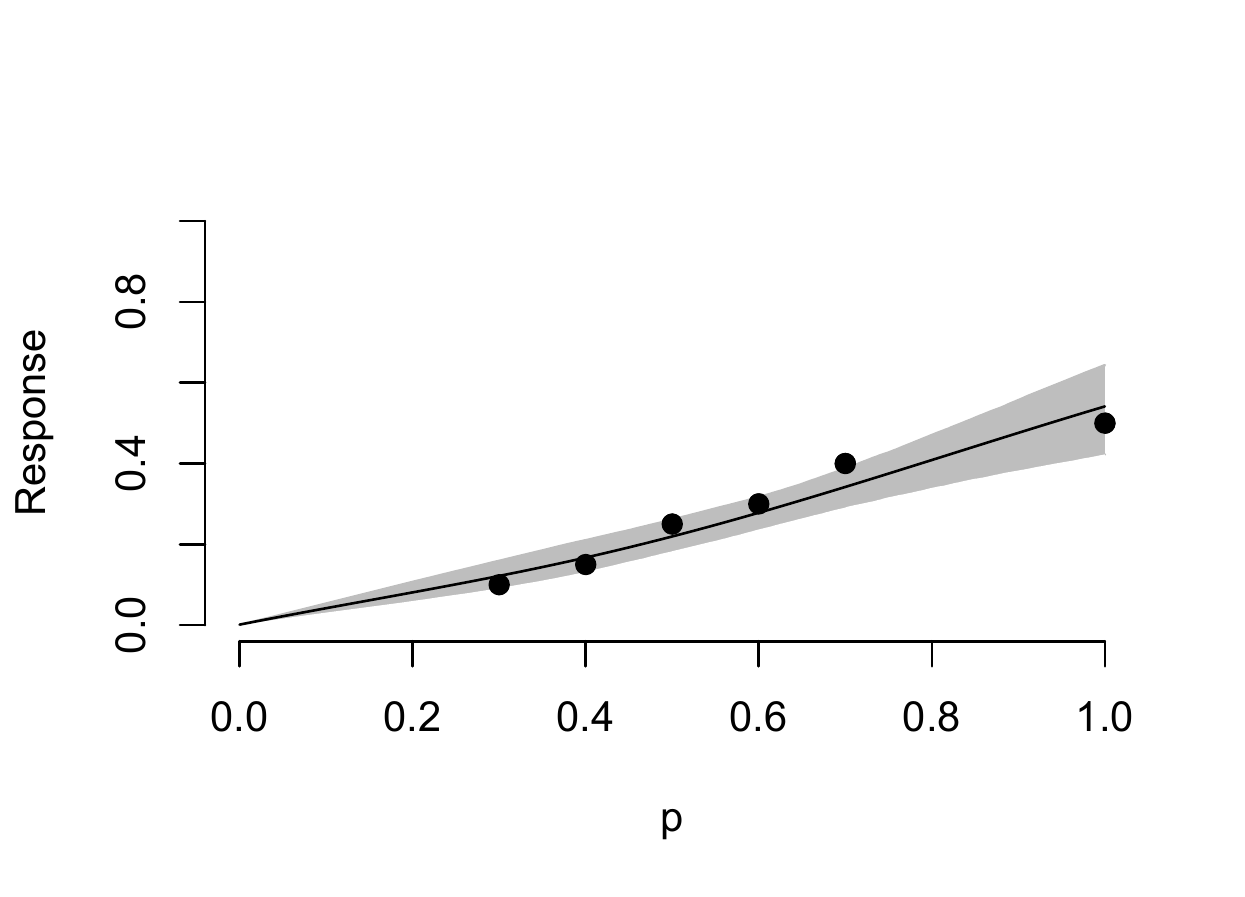}
\includegraphics[width=3.2in]{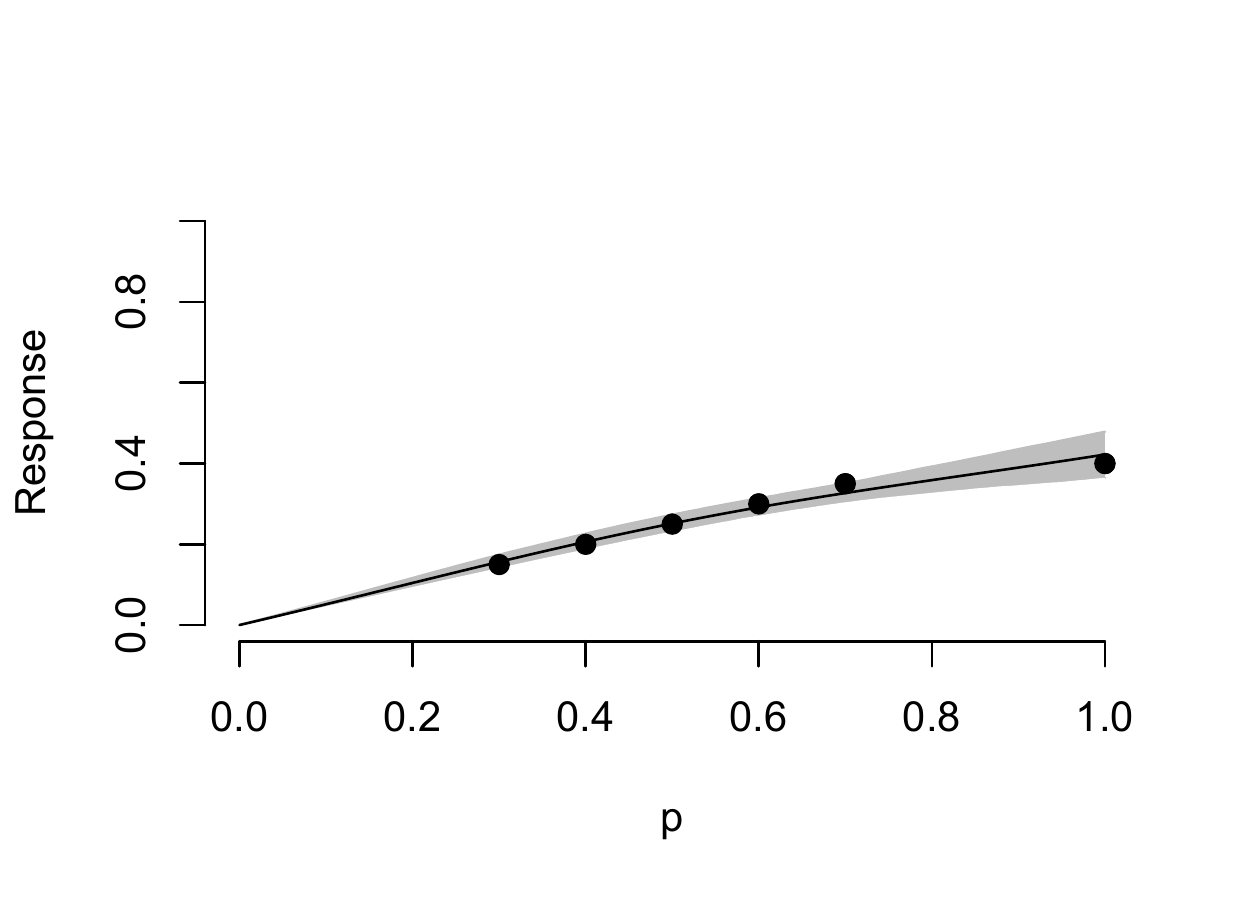}
\includegraphics[width=3.2in]{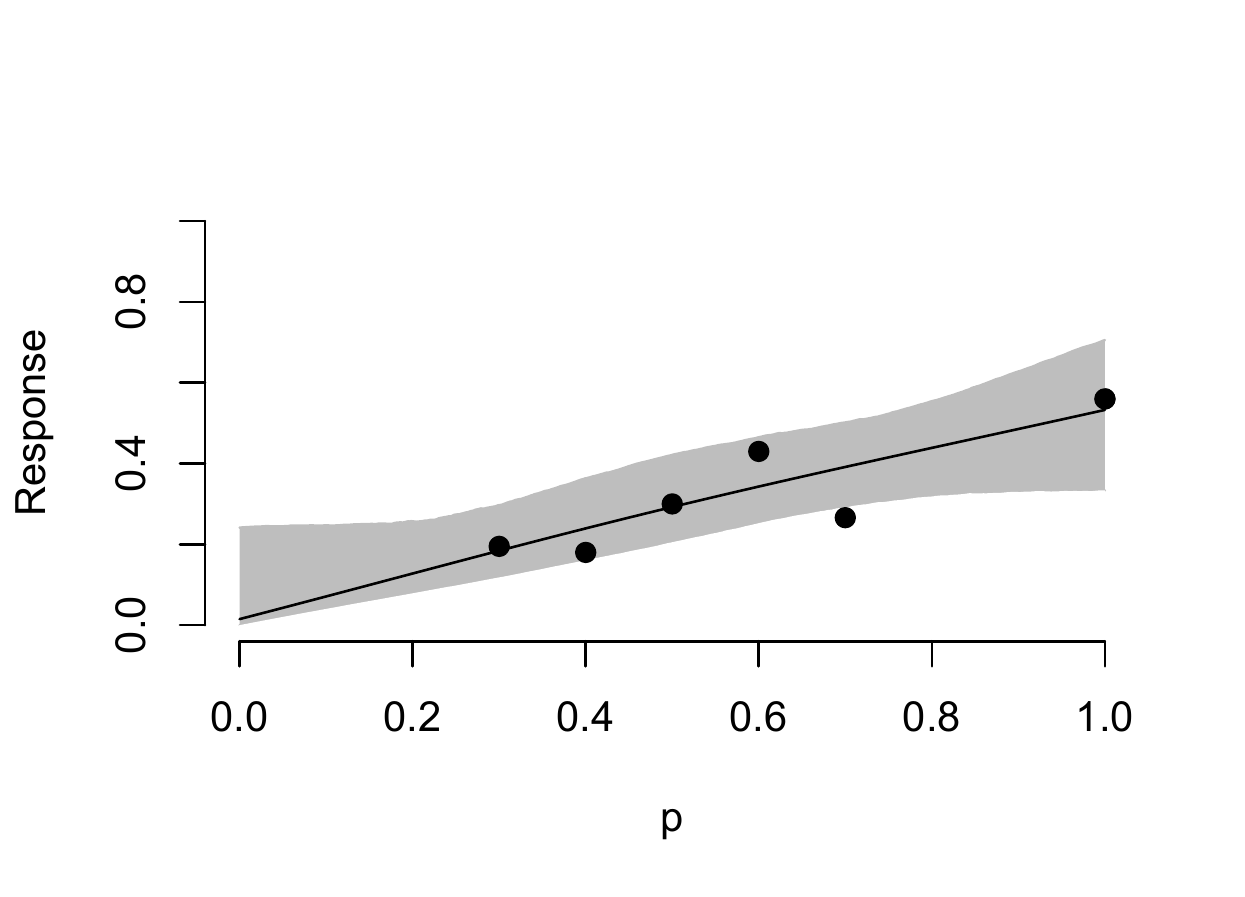}
\includegraphics[width=3.2in]{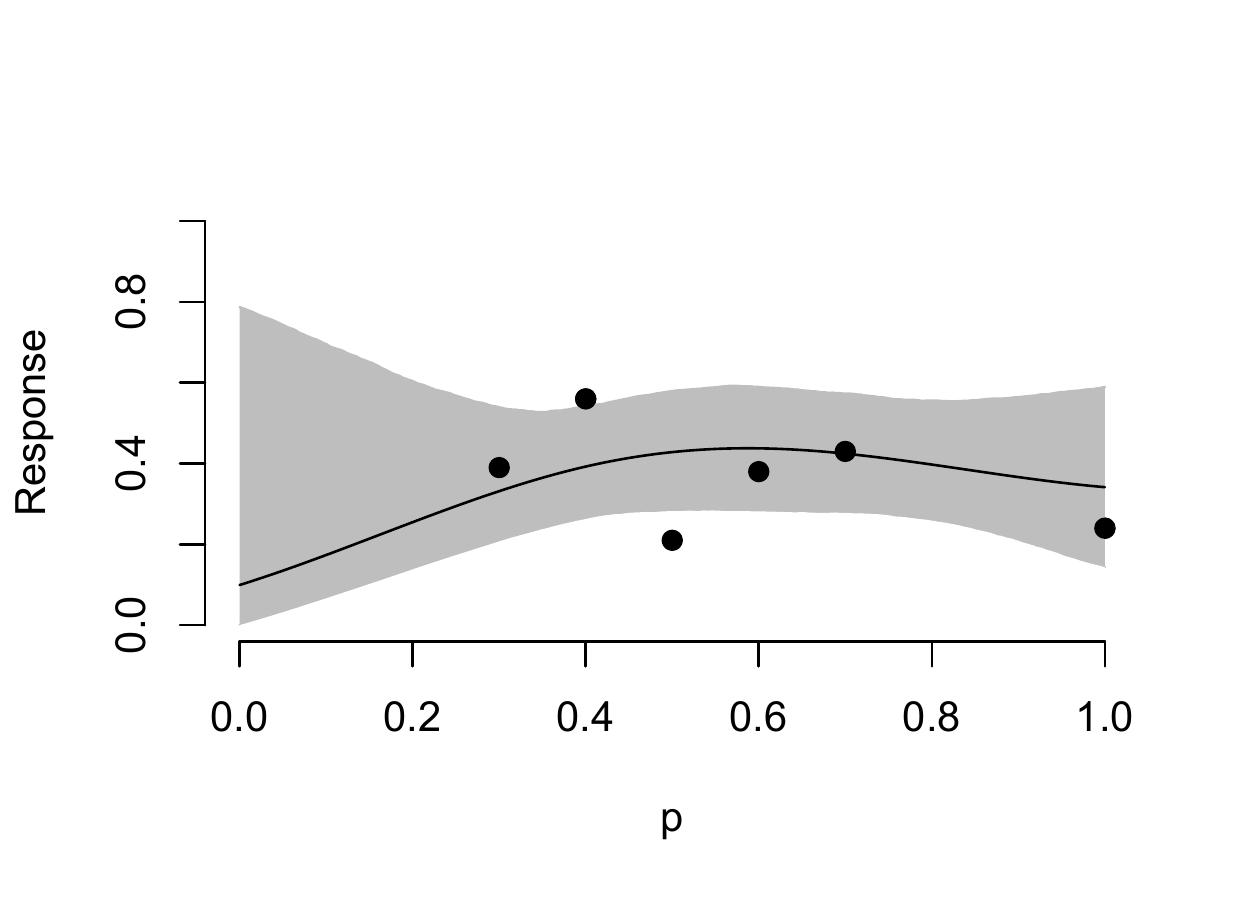}
\end{center}
\caption{Posterior summaries for four representative subjects.  The gray shaded region reflects 95\% credible regions for the curve. The posterior mean strategy is shown in red.  The posterior probability of $k$-step compliance is (from left to right, top to bottom):  92\%, 1.4\%, 32\% and 0\%.}\label{post_curves}
\end{figure}

We also obtain, for each individual, a strategy curve estimate and corresponding point-wise 95\% posterior credible intervals, as well as an estimate that her strategy is $k$-step compatible.  An illustration is provided in Figure \ref{post_curves}, using the same subjects depicted in Figure \ref{data_ex}.  

These plots guide our intuition as to how the statistical model ``interprets" the data.  For instance, the plots reinforce the findings described above based on the posterior distributions of $\rho$ and $q_0$:   many players fail the convexity criterion, despite exhibiting responses that generally increase in $p$. We also observe that players with ``orderly" responses that fail convexity are judged to have lower probability of $k$-step compliance than players who play more haphazardly; this reflects the fact that for the imprecise player we cannot be certain that she does not intend to play convex increasing and through the origin, but simply does so poorly.  This effect can be seen in the second and third panels of Figure \ref{post_curves}.  Although the general trend of each set of responses appears roughly the same, the tighter precision of the points in the upper right panel (the first five data points lie perfectly on a line) leads to less uncertainty, as seen in the narrower shaded credible region compared to the lower left panel. This tighter uncertainty, along with a data point at $p=1$ that is better fit by a concave strategy, leads to a much lower estimated probability of $k$-step compliance (only 1.4\% compared to 32\%).

\begin{figure}
\begin{center}
\includegraphics[width=5.5in]{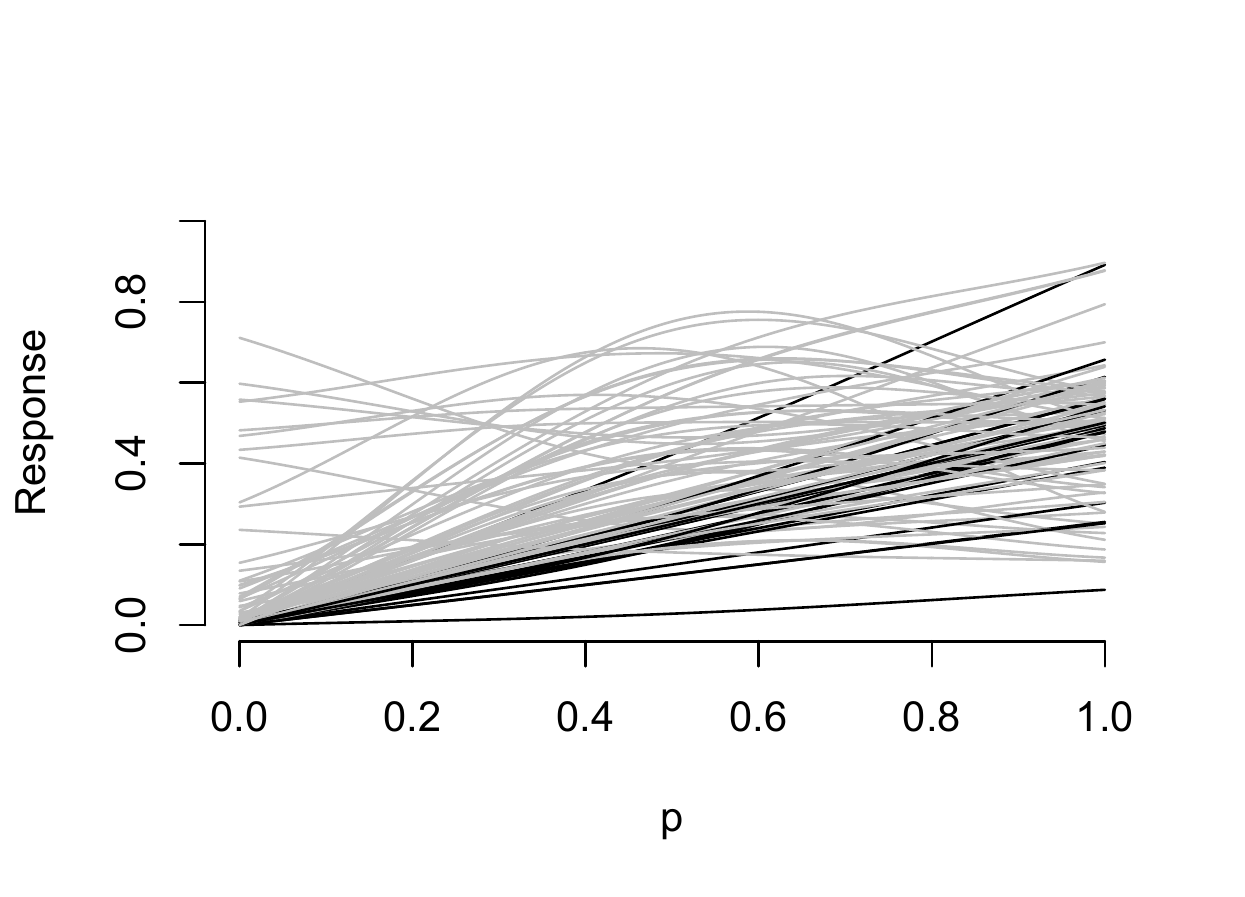}
\caption{Posterior point estimates of each player's strategy for all 106 participants.  Black curves correspond to the 23 players whose posterior probability of $k$-step compliance is greater than 1/2. Compare to Figure \ref{prior}, top panel, which displays the equivalent plot for draws from the prior.}\label{postdraws}
\end{center}
\end{figure}

Figure \ref{postdraws} shows posterior point estimates of each player's strategy.  Twenty-three players are classified as being $k$-step compliant, meaning that their posterior probability of satisfying the $k$-step compatibility criteria is greater than 1/2.  This prevalence, 23 out of 106, is only a bit less than the point estimate for $\rho q_0$ of 26\%.  The discrepancy is appropriate for two reasons.  First, note that in Figure \ref{postdraws} many of the curves are nearly linear, putting them right on the cusp of convexity.  In fact, nearly a quarter of the players have posterior probability of $k$-step compliance between 0.4 and 0.6.  This equivocal data, combined with a prior which favors the $k$-step hypothesis (i.e., $\E(\rho q_0) = 56\%$), leads to the somewhat higher posterior estimate of $\rho q_0$ than the individual estimates would suggest.

%We can also examine how strongly convex or concave each player's strategy appears to be, by visualizing the posterior distributions over $\delta_i \equiv \nu  -  (\mu^0 - \phi)\eta + \phi$.  The quantity $\delta_i$ will fall between $-1$ and $1$, with larger values indicating highly non-convex (concave) strategies and large negative values indicating highly concave strategies.  Figure \ref{delta} shows the posterior distributions for all $\delta_i$.  We see immediately that many players' data do not strongly support nor contradict convexity, as evidenced by the many posterior densities straddling zero.  A handful of players exhibit strongly convex or concave strategies.

%\begin{figure}
%\begin{center}
%\includegraphics[width=5.25in]{slides/delta.pdf}
%\end{center}
%\caption{The quantity $\delta \equiv \nu  -  (\mu^0 - \phi)\eta + \phi $ measures the degree of convexity or concavity of a strategy curve on a scale from $-1$ to $1$.  This plot depicts the posterior distributions of $\delta$ for all players.}\label{delta}
%\end{figure}

\subsection{Alternative strategy types}\label{future_work}
A promising avenue for future study would be to explicitly include plausible strategies that are not necessarily representable in the $(\mu^0, \Omega)$ structural model.  For example, we observed that several participants appear to use a ``constant increment" strategy, increasing their response by a fixed amount as $p$ increased across the experimental values $\lbrace 0.3,0.4,0.5,0.6,0.7,1\rbrace$.  See, for example, the data in the upper right panel of Figure \ref{data_ex} (or Figure \ref{post_curves}).  Because the difference between $p=0.7$ and $p=1$ is an increase of 0.3, whereas all other consecutive values differ by 0.1, this strategy leads to a concave strategy in $p$.  On the one hand, our statistical model correctly classifies this as a non-$k$-step strategy.  On the other hand, the Beta noise specification entails that this strategy is still given an 1.4\% chance of being $k$-step compliant, whereas in light of the alternative ``constant increment" hypothesis this estimate seems much to lenient.  Enriching the model with such specific alternative hypotheses would refine the conclusions of the analysis.

Some subjects' responses strongly suggest the influence of priming effects \citep{tversky1981framing}, whereby game responses are systematically affected by cues provided in the instruction set. Specifically, many players copied the demonstrated game play at values $p=0.3$ and $p=0.4$ exactly. The ``constant increment" strategy described above seems to be a byproduct of naively extrapolating from these two seed values.  Because all players were given the same instructions, this effect should not interfere with our main conclusion about the prevalence of $k$-step thinking, but does undermine the idea that such players are using $k$-step thinking even if their responses happened to be convex, increasing and through the origin.  Future experimental designs could leverage the presence of priming effects to better ascertain which players use $k$-step thinking, by introducing cues that, if attended to, could not possibly result in $k$-step compatible strategy.  Modified instructions could, for example,  include example strategies that are non-monotonic functions of $p$.

\subsection{Covariate analysis}
To investigate which player attributes are predictive of playing a $k$-step compliant strategy, we regress the posterior probability of being a $k$-step thinker on the player attributes shown in Table \ref{correlation_table}.  Only time-taken and the cognitive reflection task indicator appear predictive of $k$-step compliance.  A more sophisticated regression-tree analysis (not reported here) yielded similar conclusions.  Our findings match those of \cite{burnham2009higher}, who find that lower responses in a $p$-beauty contest are positively associated with a cognitive reflection task and negatively associated with age.

\begin{table}[h!]
\begin{center}
\begin{tabular}{lrl}\label{correlation_table}
Variable name & estimate (\%points) & $p$-value\\
\midrule
time taken on games (minutes) & 3.9&{\bf 0.005} \\
gender (male)&1.9&0.71\\
education (bachelors)&-4.1&0.69\\
education (graduate degree)&11.4&0.34\\
education (high school graduate)&3.2&0.81\\
education (some college)&0.3&0.97\\
education (some high school)& 2.2&0.77 \\
age in years& -0.23& 0.38 \\
cognitive reflection task indicator&20.3&{\bf 0.0006}
 \end{tabular}\caption{Regressing the posterior probability of $k$-step compliance against the above attributes reveals that the cognitive reflection dummy variable is the dominant predictor.}
\end{center}
\end{table}
In future work, factors that might more strongly predict $k$-step compliance could be included.  For example, while not as fine-grained as fMRI measurements, recording the number of times a subject edits his web form may be an excellent proxy of deliberative reasoning.  Ideally, such external confirmatory evidence would be incorporated directly within the hierarchical model, rather than being regressed upon ex post.  In a model expanded to include proxies for deliberative thinking, subjects would be grouped by these additional measurements, and pooling of data (via the hierarchical structure of the model) would occur only between similar subjects.

In light of our finding that many people do not apply $k$-step reasoning, a future line of research would be to conduct experiments to determine various ways that individuals might be induced to do so.  An individual's posterior probability of $k$-step reasoning, as produced from our analysis, could serve as an outcome variable in experiments that vary incentives, instructions, and practice on similar games.  

\section{Summary}
Understanding how people reason about the reasoning of others---how they think strategically in multiplayer games---is a fundamental question of behavioral science. Experimental evidence from simple multiplayer games has revealed that real players do not adhere to the theoretically optimal strategies defined by classical game theory.  Alternative, non-equilibrium, theories based on iterated reasoning have been shown to be better at predicting how players will actually respond in such games.  However, the sheer diversity of behavior observed in beauty contest data makes model estimation difficult.  Models that are sufficiently flexible are unidentified and unilluminating, while estimates from more restricted models are liable to be distorted by data from subjects who violate the assumptions of the given model.

We contribute to this literature by analyzing data from a newly conducted beauty contest experiment.  Our analysis proceeds in two steps.  First, we introduce a flexible $k$-step thinking model to describe player responses in beauty contest experiments and deduce two compatibility criteria that strategies generated according to this model must have. Second, we fit an even more flexible, purely descriptive statistical model to the player response data.  We identify $k$-step compatible strategies by observing which players' strategies satisfy these compatibility criteria.  This approach allows (an upper bound on) the prevalence of $k$-step strategies to be estimated without imposing assumptions about the distribution of player types or on players' beliefs about this distribution.   

Using our approach, we estimate that approximately 25\% of participants from our study population (Amazon Turk users in the United States) play $k$-step compatible strategies.  This figure is somewhat lower, but broadly consistent with, the comparable study of \cite{coricelli2009neural} which was based on a much smaller sample size.

An interesting implication of our analysis is that additional research regarding the $k$-step {\em in}compatible strategies could prove fruitful.  Many of the inferred strategies, while demonstrably not $k$-step compatible, appear far more structured than would be expected from a purely random player. Careful examination of our individual-level strategy curve estimates suggests that more elaborate error models --- such as non-strategic models based on known cognitive biases, such as priming --- could help refine future estimates of the prevalence of true strategic thinkers.  Individual-level estimates from hierarchical Bayesian models, such as ours, can also be used as outcome variables in experiments designed to investigate determinants of strategic sophistication.  We hope that our study will motivate additional work along these lines and prove a useful tool for future empirical studies on strategic reasoning in games.

\bibliographystyle{abbrvnat}

\bibliography{CHbib}
\section*{Appendix A:  Computational details}
Posterior sampling is performed via a Metropolis-within-Gibbs approach.  We sequentially sample parameters, given the current value of all other parameters, according to the Metropolis-Hastings acceptance probabilities.  A challenge with this approach is to devise suitable proposal densities.  We utilize a random-walk approach.  As many of the model parameters are restricted to the unit interval, we introduce a latent variable representation using the ``wrapping function":
$$g(x) = x - \lfloor x \rfloor + \mathbb{1}(x < 0),$$
where $\lfloor x \rfloor$ denotes the integer part of $x$.  This function maps numbers in the unit interval to themselves, while numbers outside the unit interval get mapped back to their fractional part, in the case of positive numbers, or one minus their fractional part, in the case of negative numbers.  So 0.8 gets mapped to itself; 1.3 gets mapped to 0.3; -2.4 gets mapped to 0.6.  For a parameter, such as $\mu_0$, restricted to the unit interval, we conduct a random walk over parameters $\tilde{\mu}$ on the whole real line and define $\mu \equiv g(\tilde{\mu})$.  The $\tilde{\mu}$ parameters are unidentified, but this is irrelevant as we report inferences and define priors on the original, identified scale.  (Note that $\eta_i$ is restricted to $[0.3, 0.7]$ rather than $[0, 1]$, so $g(\cdot)$ must be modified accordingly by a simple rescaling.)

For the $\phi_i$ parameters, which can be exactly zero with positive probability, we also introduce a binary latent variable $z_i$ and define $\phi_i = z_i g(\tilde{\phi_i})$.  We conduct the random walk on the real line with $\tilde{\phi}$ and transform to $\phi_i$ for likelihood evaluations.

\begin{enumerate}
\item For each $i$, sample ($\eta_i, \phi_i, \nu_i, \mu_i \mid \mbox{--}$) according to a Metropolis ratio using likelihood and prior given in (\ref{likelihood}) and  (\ref{prior}) respectively.  Proposal density evaluations can be avoided by using a symmetric random walk over the elements of $\tilde{\eta}_i, \tilde{\phi}_i, \tilde{\nu}_i$ and $\tilde{\mu}^0_i$, centered at the current values.
\item For each $i$, sample ($z_i \mid \mbox{--}$) using a straightforward application of Bayes rule.
\item For each $i$, sample ($s_i \mid \mbox{--}$) using a straightforward application of Bayes rule.
\item Sample ($\bf{w} \mid \mbox{--}$) from a Dirichlet distribution with parameter $\alpha^* = \boldsymbol{\alpha} + \boldsymbol{\kappa}$, where $\boldsymbol{\kappa}$ records counts of how many observations are currently assigned to each level of $s$.  
\item Sample ($\rho \mid \mbox{--}$) as a Beta($\gamma$, $\beta$) random variable.  Let $n_1 = \sum_i z_i$ and $n_0 = \sum_i (1 - z_i)$.  Then $\gamma = 3 + n_0$ and $\beta = 1 + n_1$.
\item Sample ($q_0 \mid \mbox{--}$) as a Beta($\gamma$, $\beta$) random variable. Let $n_{q0}$ be the number of players for whom both $\phi_i =0$ and $\nu_i < \mu_i^0 \eta_i$.  Then  $\gamma = 3 + n_{q0}$ and $\beta = 1 + n_0 - n_{q0}$.
\item Sample ($q_1 \mid \mbox{--}$) s a Beta($\gamma$, $\beta$) random variable. Let $n_{q1}$ be the number of players for whom both $\phi_i \neq 0$ and $\nu_i < \mu_i^0 \eta_i$.  Then  $\gamma = 3 + n_{q1}$ and $\beta = 1 + n_0 - n_{q1}$.

\end{enumerate} 

\section*{Appendix B:  Diagnostic plots}
Below, we include trace plots for key parameters from the MCMC sampler. By this rough metric, mixing looks adequate. Subject-specific samples are shown, arbitrarily, for participant \#100.  Although hard to discern in the plot for $\phi$, 75\% of the samples drawn are $\phi = 0$. Also we provide a Geweke-Brooks diagnostic plot for $\rho$, $q_0$ and $q_1$; for an explanation of this plot, see \cite{coda}.

\begin{center}
\includegraphics[width=5in]{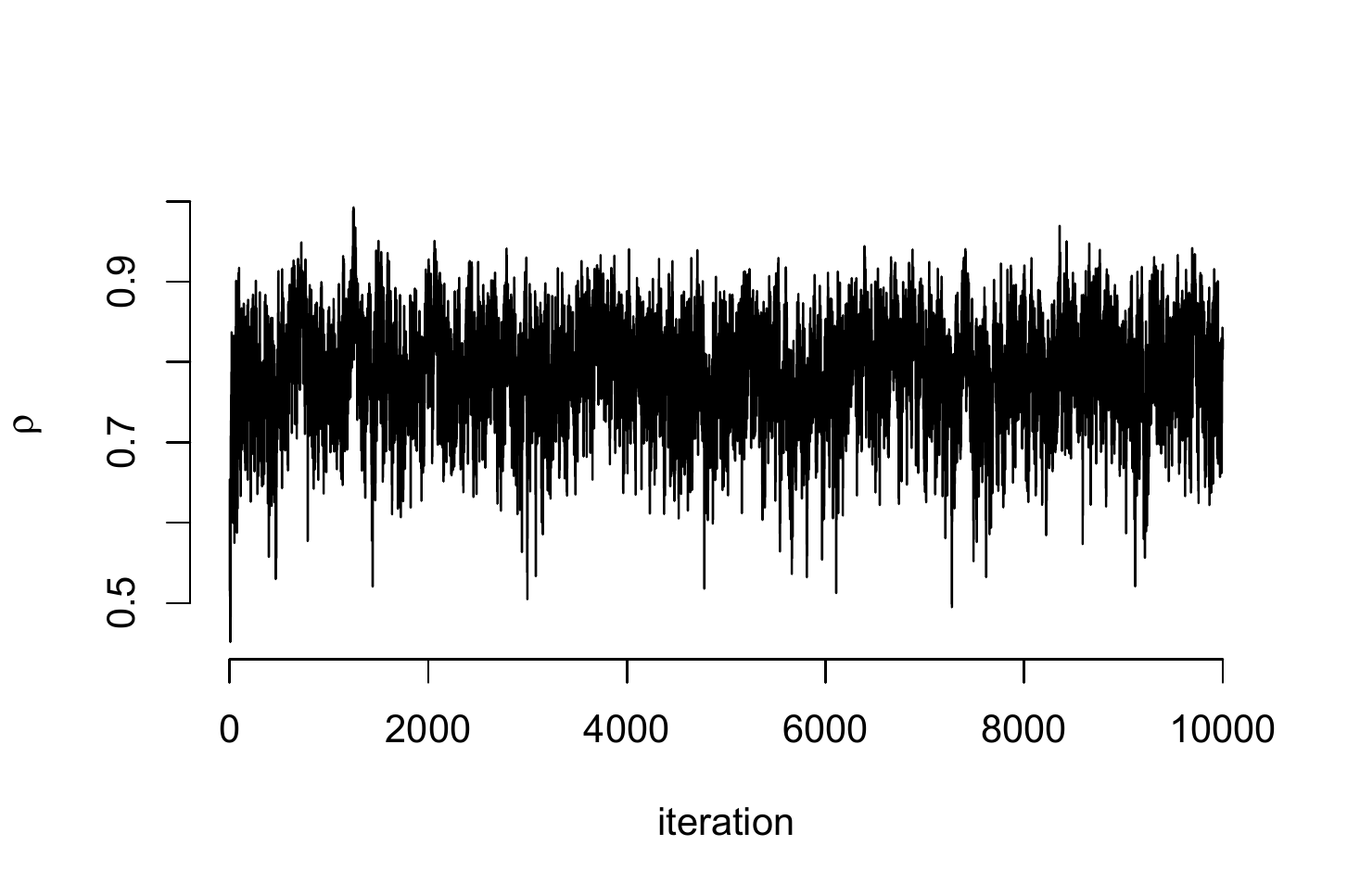}\\
\includegraphics[width=5in]{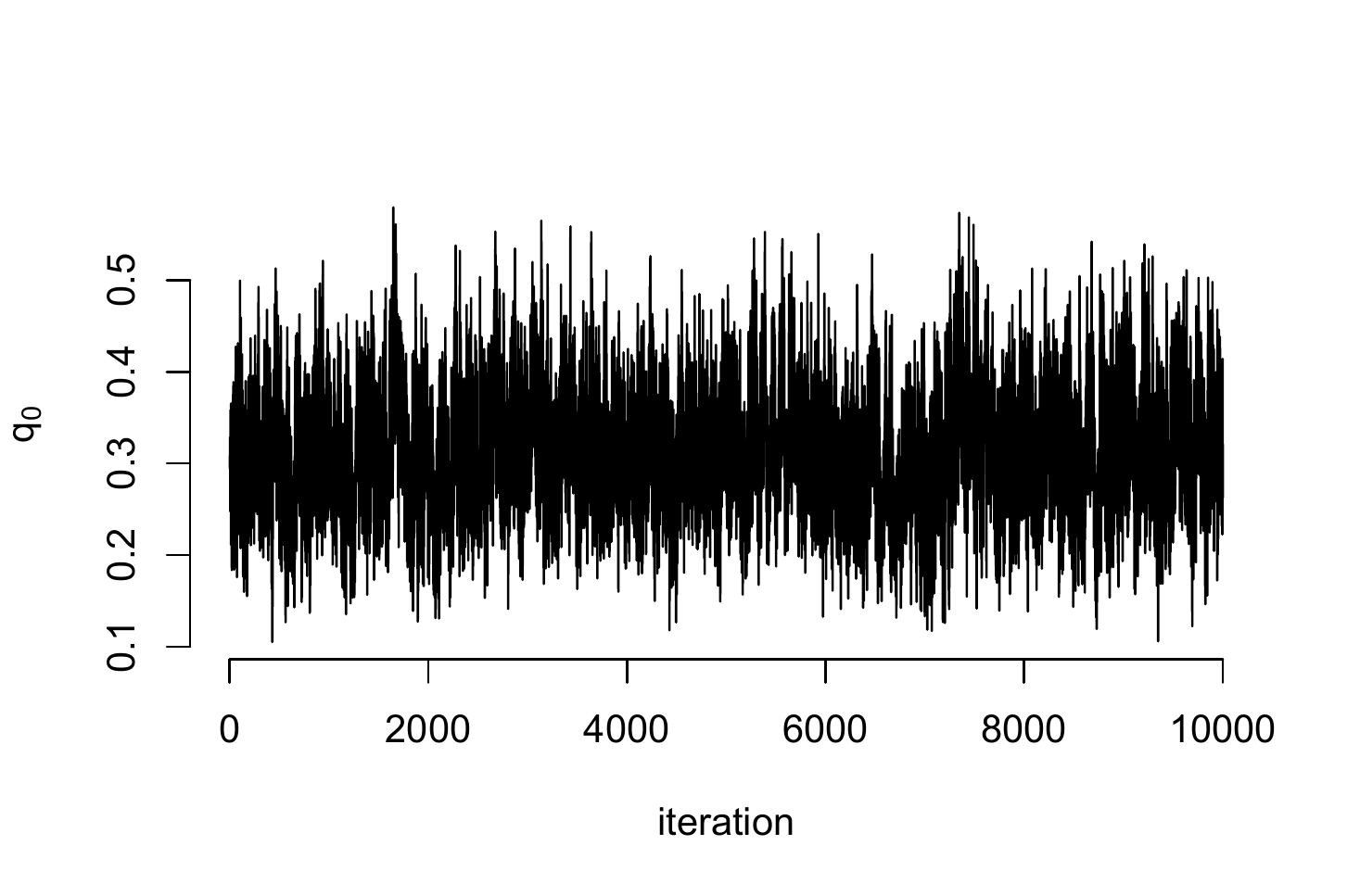}
\includegraphics[width=5in]{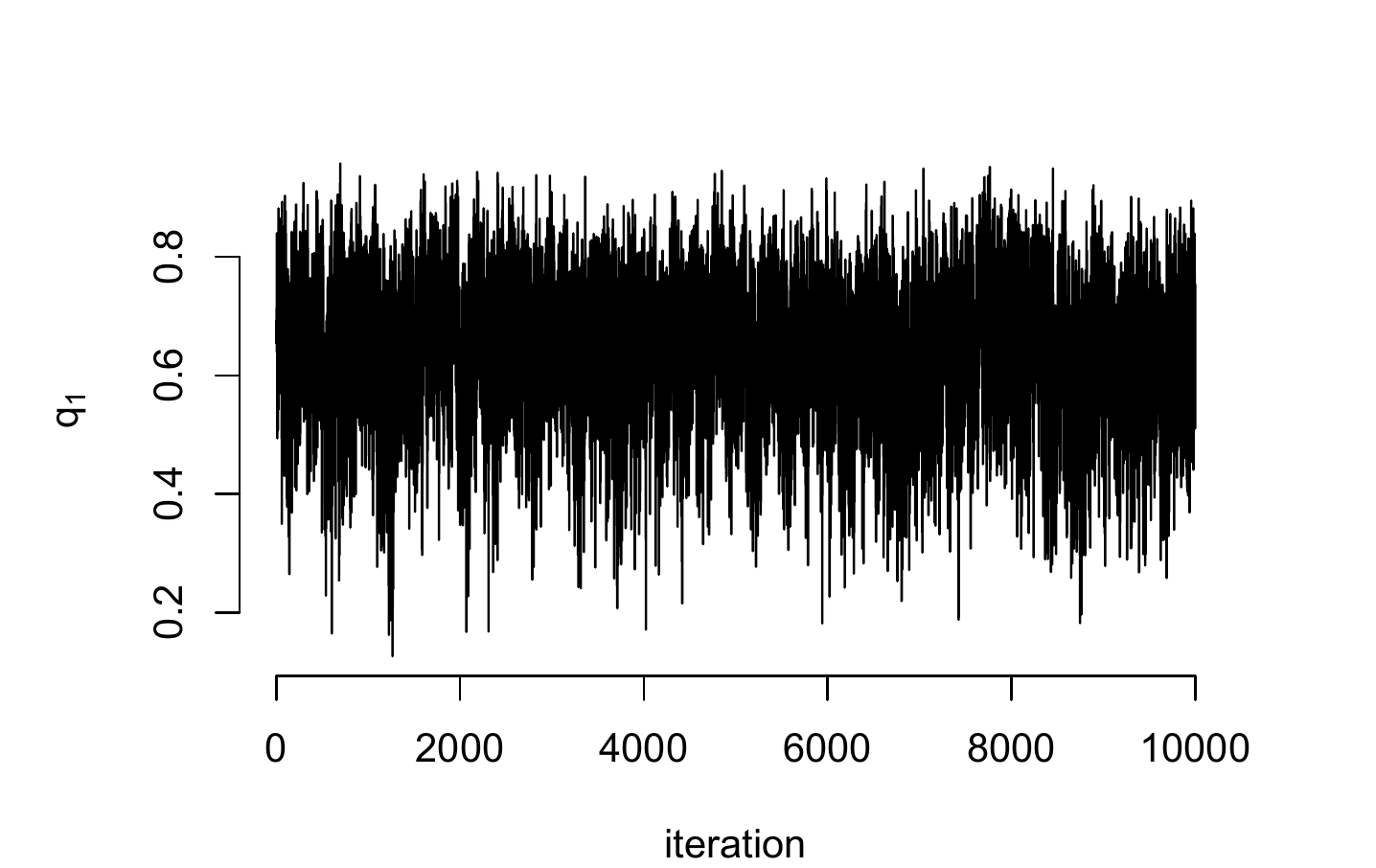}
\includegraphics[width=5in]{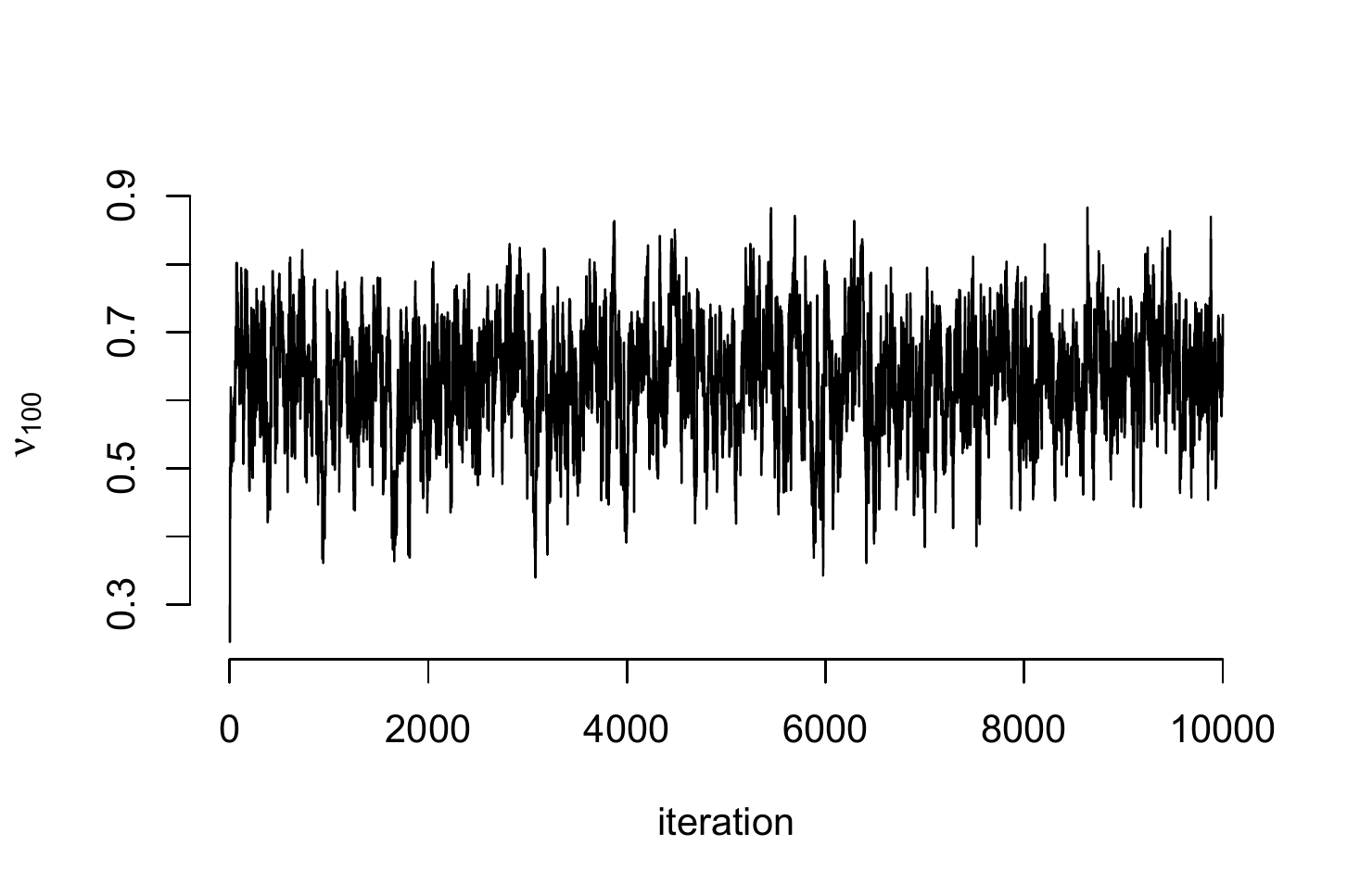}
\includegraphics[width=5in]{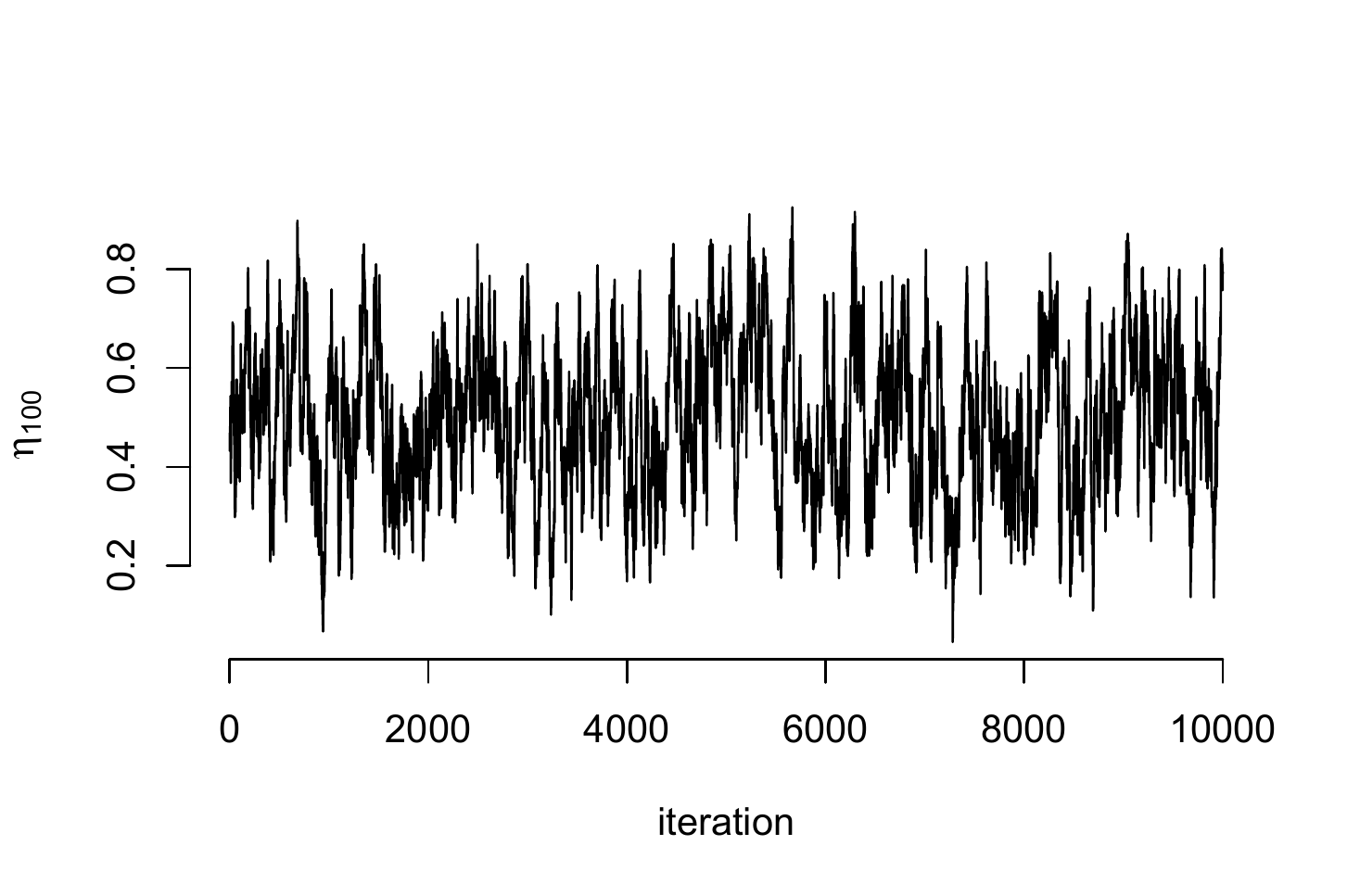}
\includegraphics[width=5in]{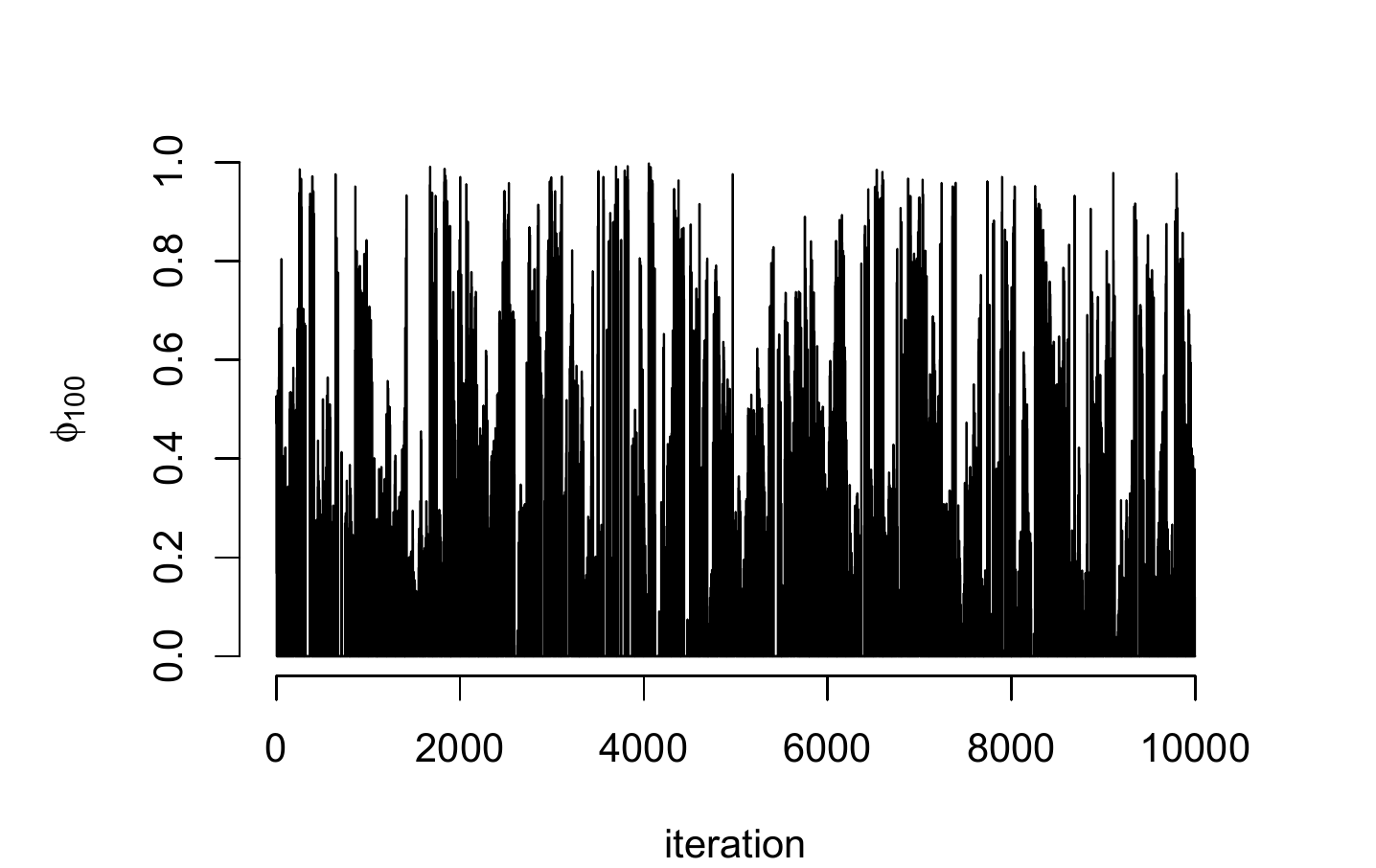}
\includegraphics[width=5in]{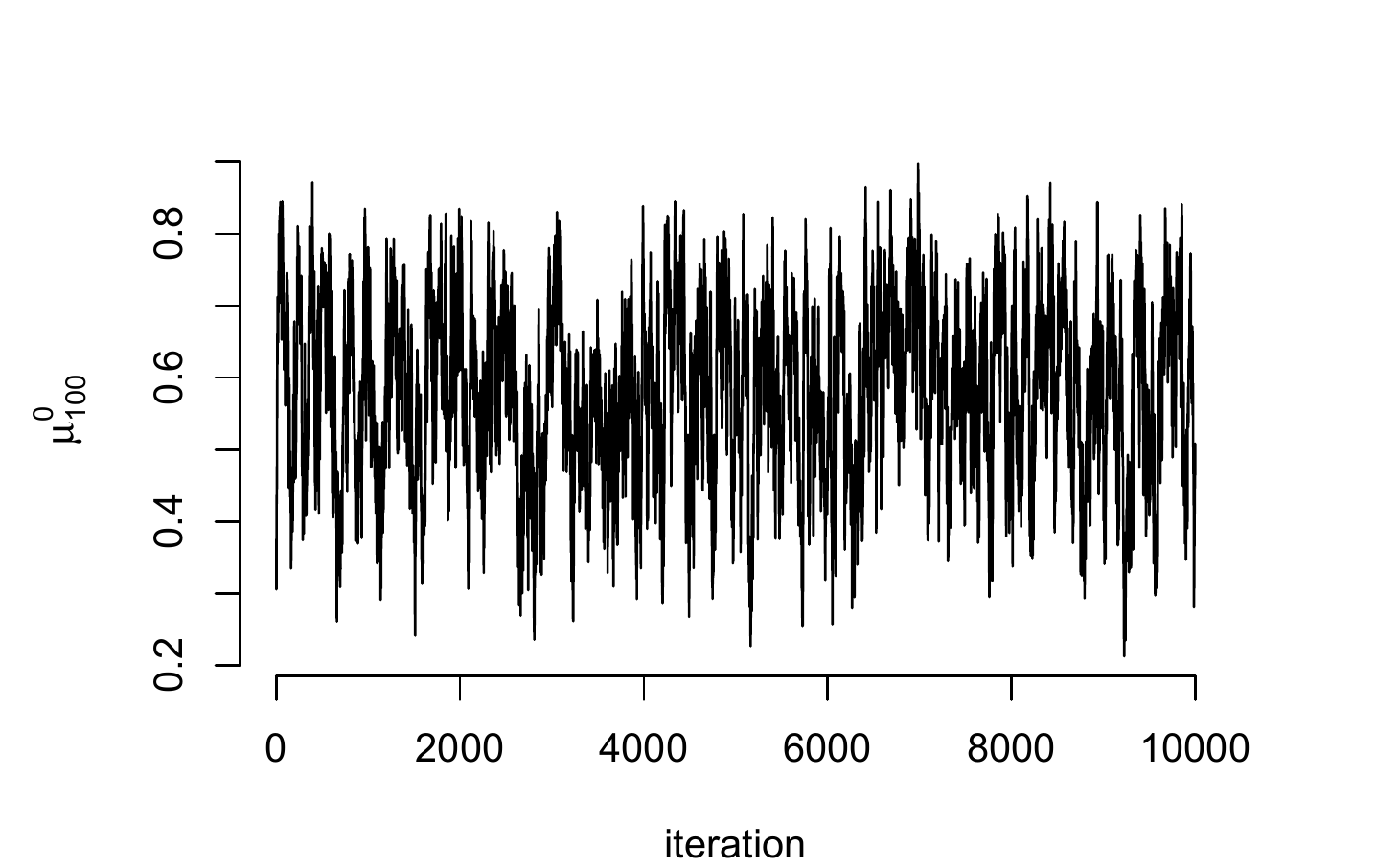}
\end{center}

\begin{center}
\includegraphics[width=5in]{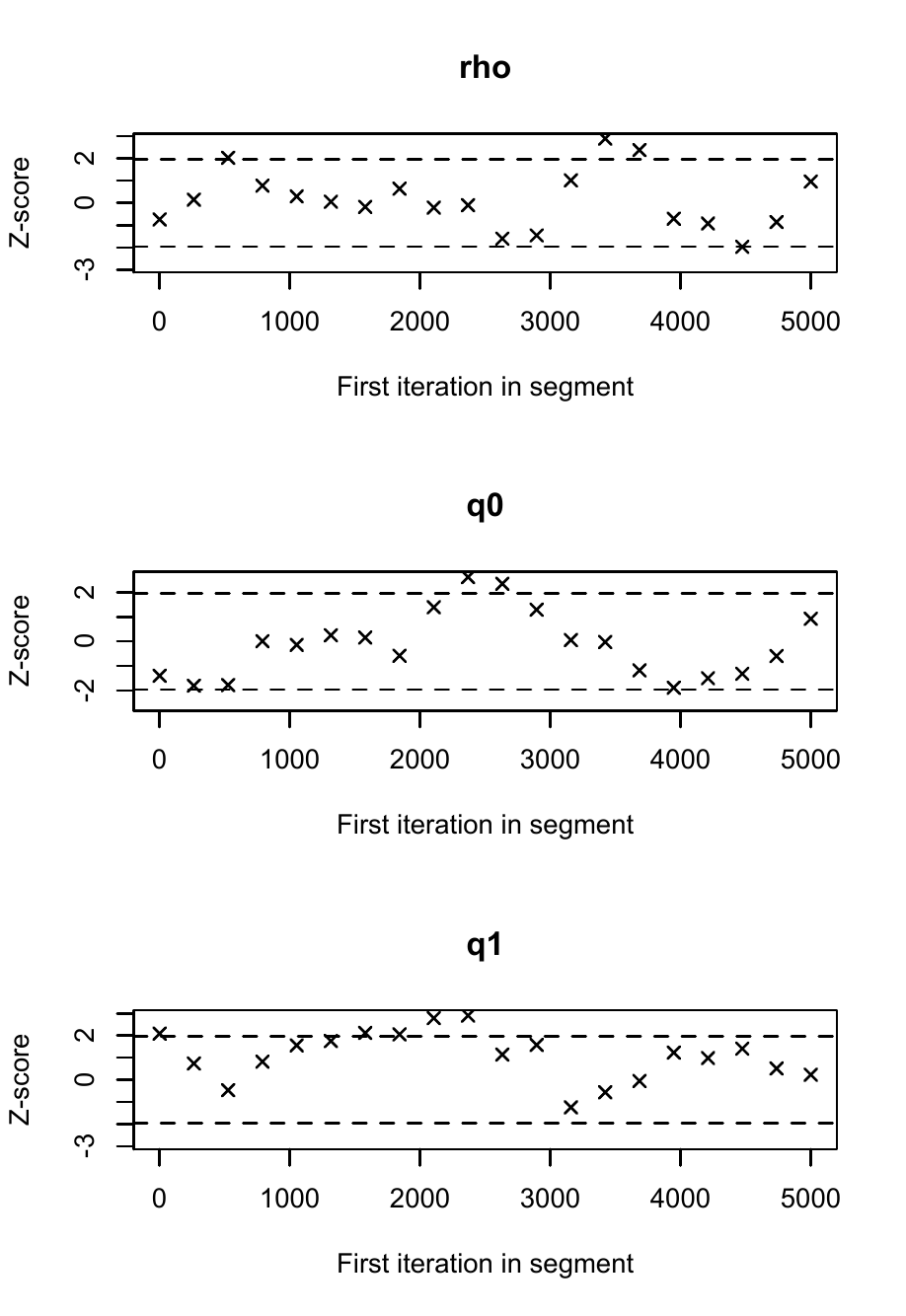}
\end{center}

\end{document}